\documentclass[12pt]{article}
\usepackage[utf8]{inputenc}
\usepackage[usenames,dvipsnames]{color}
\usepackage{amsmath}
\usepackage{amssymb}
\usepackage{mathptmx}
\usepackage{authblk}
\usepackage{ulem}
\usepackage{comment}

\usepackage[a4paper, margin=1in]{geometry}
\usepackage{multirow}
\usepackage{array}
\usepackage{rotating}
\usepackage{booktabs}%
\usepackage{float}

\usepackage[breaklinks,colorlinks,citecolor=black,linkcolor=black]{hyperref}
\usepackage[comma,sectionbib,numbers, round,sort&compress]{natbib}
\usepackage{graphicx}
\usepackage{subcaption}
\usepackage[labelfont=bf]{caption}
\DeclareCaptionLabelSeparator{bar}{ \textbar{} }
\newcommand{\arcsec}{^{\prime\prime}}

\newcommand{\target}{IRAS 18134-1942 }

\def\jnl@style{\it}
\def\aaref@jnl#1{{\jnl@style#1}}

\newcommand{\mycite}[1]{\textit{\cite{#1}}}

\renewcommand{\textcolor}[2]{#2}

    \includecomment{exfig}

\usepackage{etoolbox}
\makeatletter
\patchcmd{\@maketitle}{\null\vskip 2em}{}{}{}
\makeatother

\newenvironment{myabstract}{%
  \list{}{%
    \leftmargin1.5em
    \rightmargin0pt}%
    \item\relax
    \small}
{\endlist}

\title{A misaligned protostellar disk fed by gas streamers in a barred spiral-like massive dense core}  

\author[1,2,3*]{Xiaofeng Mai}
\author[1,4*]{Tie Liu}
\author[1,5]{Xunchuan Liu}
\author[1,4]{Bo Zhang}
\author[6]{Paul F. Goldsmith}
\author[7]{Neal J. Evans II}
\author[8]{Qizhou Zhang}
\author[9,10]{Kee-Tae Kim}
\author[11]{Dongting Yang}
\author[3]{Mika Juvela}
\author[12,13,14]{Fengwei Xu}
\author[1]{Wenyu Jiao}
\author[11]{Hongli Liu}
\author[15]{Patricio Sanhueza}
\author[16,17]{Guido Garay}
\author[18]{Xi Chen}
\author[11]{Shengli Qin}
\author[3]{Jakobus M. Vorster}
\author[19]{Anandmayee Tej}
\author[20]{Zhiyuan Ren}
\author[21]{Sami Dib}
\author[22,23]{Shanghuo Li}
\author[1,2,24]{Qiuyi Luo}
\author[25,26]{Jihye Hwang}
\author[27,28]{Prasanta Gorai}
\author[29]{Ariful Hoque}
\author[30]{Yichen Zhang}
\author[31,32]{Jeong-Eun Lee}
\author[16]{Siju Zhang}
\author[3]{Emma Mannfors}
\author[3]{Devika Tharakkal}
\author[33]{Lokesh Dewangan}
\author[16]{Leonardo Bronfman}
\author[17,34]{Pablo García}
\author[35]{Xindi Tang}
\author[16]{Swagat R. Das}
\author[35]{Gang Wu}
\author[9]{Chang-Won Lee}
\author[36,37,38]{James O. Chibueze}
\author[1]{Yankun Zhang}
\author[1]{Qilao Gu}
\author[24,39]{Ken’ichi Tatematsu}
\author[1]{Guangli Wang}
\author[17]{Lei Zhu}
\author[1,4]{Zhiqiang Shen}

\affil[1]{Shanghai Astronomical Observatory, Chinese Academy of Sciences, Shanghai 200030, PR China; \url{maixf@shao.ac.cn}; \url{liutie@shao.ac.cn};}
\affil[2]{School of Astronomy and Space Sciences, University of Chinese Academy of Sciences, No. 19A Yuquan Road, Beijing 100049, People’s Republic of China}
\affil[3]{Department of Physics, PO Box 64, 00014, University of Helsinki, Finland}
\affil[4]{State Key Laboratory of Radio Astronomy and Technology, A20 Datun Road, Chaoyang District, Beijing, 100101, P. R. China}
\affil[5]{Leiden Observatory, Leiden University, P.O. Box 9513, 2300RA Leiden, The Netherlands}
\affil[6]{Jet Propulsion Laboratory, California Institute of Technology, 4800 Oak Grove Drive, Pasadena, CA 91109, USA}
\affil[7]{Department of Astronomy, The University of Texas at Austin, 2515 Speedway, Austin, TX 78712, USA}
\affil[8]{Center for Astrophysics | Harvard \& Smithsonian, 60 Garden Street, Cambridge, MA 02138, USA}
\affil[9]{Korea Astronomy and Space Science Institute (KASI), 776 Daedeokdae-ro, Yuseong-gu, Daejeon 34055, Republic of Korea}
\affil[10]{University of Science and Technology, Korea (UST), 217 Gajeong-ro, Yuseong-gu, Daejeon 34113, Republic of Korea}
\affil[11]{School of Physics and Astronomy, Yunnan University, Kunming, 650091, People's Republic of China}
\affil[12]{Department of Astronomy, Peking University, 100871 Beijing, People’s Republic of China}
\affil[13]{Kavli Institute for Astronomy and Astrophysics, Peking University, 5 Yiheyuan Road, Haidian District, Beijing 100871, People’s Republic of China}
\affil[14]{I. Physikalisches Institut, Universität zu Köln, Zülpicher Straße 77, 50937 Cologne, Germany}
\affil[15]{Department of Astronomy, School of Science, The University of Tokyo, 7-3-1 Hongo, Bunkyo, Tokyo 113-0033, Japan}
\affil[16]{Departamento de Astronomía, Universidad de Chile, Camino el Observatorio 1515, Las Condes, Santiago, Chile}
\affil[17]{Chinese Academy of Sciences South America Center for Astronomy, National Astronomical Observatories, Chinese Academy of Sciences, Beijing 100101, China}
\affil[18]{Center for Astrophysics, Guangzhou University, Guangzhou 510006, People’s Republic of China}
\affil[19]{Indian Institute of Space Science and Technology, Thiruvananthapuram 695 547, Kerala, India}
\affil[20]{National Astronomical Observatories, CAS, Beijing, China}
\affil[21]{Max-Planck-Institut f\"{u}r Astronomie, K\"{o}nigstuhl 17, D-69117 Heidelberg, Germany}
\affil[22]{School of Astronomy and Space Science, Nanjing University, 163 Xianlin Avenue, Nanjing 210023, People’s Republic of China}
\affil[23]{Key Laboratory of Modern Astronomy and Astrophysics (Nanjing University), Ministry of Education, Nanjing 210023, People’s Republic of China}
\affil[24]{National Astronomical Observatory of Japan, National Institutes of Natural Sciences, 2-21-1 Osawa, Mitaka, Tokyo 181-8588, Japan}
\affil[25]{Institute for Advanced Study, Kyushu University, 774 Motooka Nishi-ku Fukuoka, Japan}
\affil[26]{Department of Earth and Planetary Sciences, Faculty of Science, Kyushu University, Nishi-ku, Fukuoka 819-0395, Japan}
\affil[27]{Rosseland Centre for Solar Physics, University of Oslo, PO Box 1029 Blindern, 0315, Oslo, Norway}
\affil[28]{Institute of Theoretical Astrophysics, University of Oslo, PO Box 1029 Blindern, 0315, Oslo, Norway}
\affil[29]{S. N. Bose National Centre for Basic Sciences, Block-JD, Sector-III, Salt Lake City, Kolkata 700106, India}
\affil[30]{Department of Astronomy, Shanghai Jiao Tong University, 800 Dongchuan Rd., Minhang, Shanghai 200240, People's Republic of China}
\affil[31]{Department of Physics and Astronomy, Seoul National University, 1 Gwanak-ro, Gwanak-gu, Seoul 08826, Republic of Korea}
\affil[32]{SNU Astronomy Research Center, Seoul National University, 1 Gwanak-ro, Gwanak-gu, Seoul 08826, Republic of Korea}
\affil[33]{Physical Research Laboratory, Navrangpura, Ahmedabad 380 009, India}
\affil[34]{Instituto de Astronom\'ia, Universidad Cat\'olica del Norte, Av. Angamos 0610, Antofagasta, Chile}
\affil[35]{Xinjiang Astronomical Observatory, Chinese Academy of Sciences (CAS), Urumqi 830011, People's Republic of China}
\affil[36]{Centre for Space Research, North-West University, Potchefstroom 2520, South Africa}
\affil[37]{Department of Mathematical Sciences, University of South Africa, Cnr Christian de Wet Rd and Pioneer Avenue, Florida Park, 1709, Roodepoort, South Africa}
\affil[38]{Department of Physics and Astronomy, Faculty of Physical Sciences, University of Nigeria, Carver Building, 1 University Road, Nsukka 410001, Nigeria}
\affil[39]{Astronomical Science Program, Graduate Institute for Advanced Studies, SOKENDAI, 2-21-1 Osawa, Mitaka, Tokyo 181-8588, Japan}

\begin{document}
\maketitle             
\begin{myabstract}

\textcolor{red}{High-mass stars, born in massive dense cores (MDCs), profoundly impact the cosmic ecosystem through feedback processes and metal enrichment, yet little is known about how MDCs assemble and transfer mass across scales to form high-mass young stellar objects (HMYSOs). Using multi-scale (40–2500 au) observations of an MDC hosting an HMYSO, we identify a coherent dynamical structure analogous to barred spiral galaxies: three $\sim$20,000 au spiral arms feed a $\sim$7,500 au central bar, which channels gas to a $\sim$2,000 au pseudodisk. Further accretion proceeds through the inner structures, including a Keplerian disk and an inner disk ($\sim$100 au), which are thought to be driving a collimated bipolar outflow. This is the first time that these multi-scale structures (spiral arms, bar, streamers, envelope, disk, and outflow) have been simultaneously observed as a physically coherent structure within an MDC. Our discovery suggests that well-organized hierarchical structures play a crucial role during the gas accretion and angular momentum build-up of a massive disk.}

\end{myabstract}

\phantomsection
\label{sec:Main body}
\noindent{\large \textbf{Introduction}}

\vspace{0.5em}
\phantomsection
\label{sec:intro}

High-mass stars are known to form within clustered environment \mycite{Tan2014,Krum2014,2018ARA&A..56...41M,2019ARA&A..57..227K} and the references therein). Unlike their low-mass solar-type siblings that form from the monolithic collapse of a well-defined pre-stellar core \mycite{1987ARA&A..25...23S, 2003ApJ...585..850M, 2011A&A...527A.135C}, the formation of the massive stars relies on multi-scale accretion from much larger ($\sim$0.1–1 pc) gas reservoirs, called massive dense cores (MDCs) \mycite{2018ARA&A..56...41M}. The MDCs themselves may also continuously accrete gas from their natal clouds through larger-scale ($\sim$1-10 pc) hierarchical gas structures such as filament-hub systems \mycite{2013A&A...555A.112P,2022MNRAS.514.6038Z, 2024A&A...686A.146Z}. 
However, the formation of gas structures within MDCs and their roles in protostellar mass growth remains elusive. In addition, due to the low number statistics \mycite{2016A&ARv..24....6B,2019NatAs...3..517Z,2022NatAs...6..837L,2017NatPh..13..276C, 2023ApJ...959L..31O,2024Natur.625...55M}, it is still under debate whether stable disks widely exist in the surroundings of HMYSOs. 
\textcolor{red}{Therefore, understanding how mass is assembled and transferred across scales to massive disks is essential to constraining the physical mechanisms that govern high-mass star formation, including angular momentum regulation and disk evolution.}

\begin{figure*}[h]
\centering
\includegraphics[width=1\linewidth]{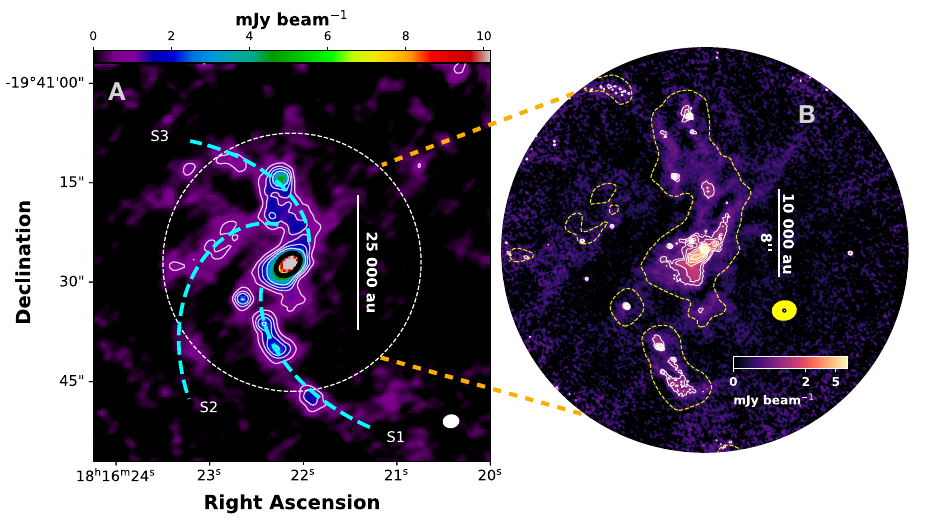}
\caption{{Continuum images of ALMA 3\,mm and 1.3\,mm observations.} {A}, ALMA 3\,mm continuum image of \target, showing a bright central core, elongated filamentary structures extending from the center, and several cores forming along those. The white contours are [3,5,7,10,15,30,50,90]$\times\sigma_{\rm 3\,mm}$, where the rms noise $\sigma_{\rm 3\,mm}$ is 0.25\,mJy per beam. The $2.15\arcsec\times 1.71\arcsec$ restored beam is plotted at the lower right corner. The cyan dashed logarithmic spiral curves are plotted for visual guidance. The white dashed circle represents the field of view of the 1.3\,mm observation. The black ellipse ($4.2\arcsec\times2.8\arcsec$ from two-dimensional Gaussian fitting) at the center represents the bar structure unresolved in 3\,mm continuum. {B,} ALMA 1.3\,mm continuum image. The yellow dashed contour marks the 3$\times\sigma_{\rm 3\,mm}$ of the 3\,mm continuum emission. The white contours are [3, 5, 9, 13, 17, 50]$\times\sigma_{\rm 1.3\,mm}$, where $\sigma_{\rm 1.3\,mm} = 0.25$ mJy per beam. The $0.31\arcsec\times 0.27\arcsec$ restored beam of the 1.3\,mm continuum image is shown as a black ellipse with black edge over the 3\,mm beam shown as a yellow ellipse between the scalebar and the colorbar. 
}
\label{cont_b3_b6}
\end{figure*}

\begin{figure*}[h]
\centering
\includegraphics[width=1\linewidth]{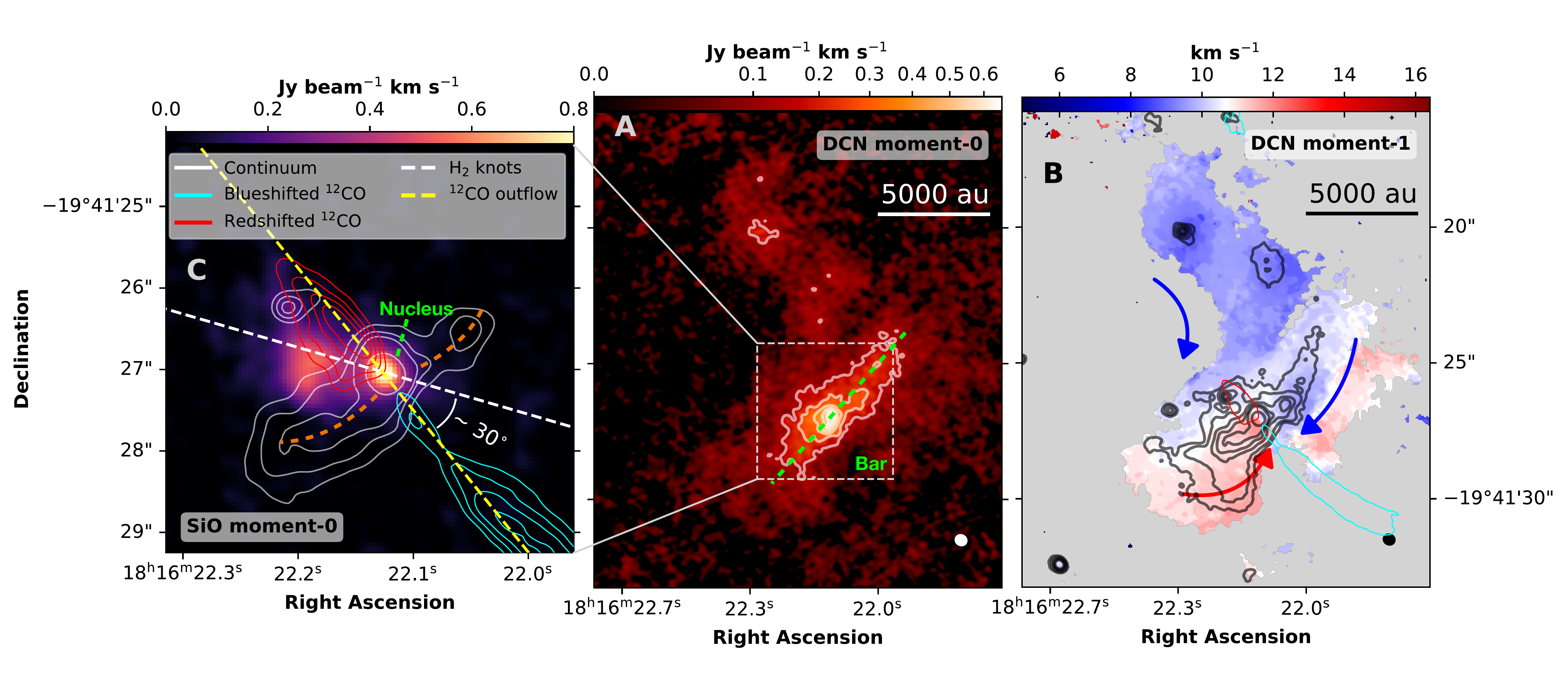}
\caption{{DCN integrated intensity and velocity maps showing infalling and converging streamers feeding the central \textcolor{red}{nucleus}, and SiO emission tracing the shocked gas and $^{12}$CO emission tracing the outflow gas.} {A} Integrated intensity of DCN. The contours are at [3, 5, 7, 9, 11]$\times\sigma_{\rm DCN}$ where $\sigma_{\rm DCN}=0.05\,{\rm Jy~beam^{-1}~km~s^{-1}}$. {B}, Centroid velocity map of DCN. The black contours are the 1.3 mm continuum emission (same as that in Fig.~\ref{cont_b3_b6}B). The blue and red arrows mark the proposed gas kinematics: three gas streamers converge to the central \textcolor{red}{nucleus}. \textcolor{red}{It is worthy noting that the northern streamer is part of the large-scale spiral arm, and the other two are consistent with the dust streamers revealed by 1.3 mm continuum emission}. The synthesized beam is shown at the lower right corners of {A} and {B}. {C}, Redshifted (red contours) and blueshifted (cyan contours) $^{12}$CO emission is overlaid of the SiO integrated intensity (colorscale). The red- and blueshifted outflow gas integrated between [20, 50] km s$^{-1}$ and [-30, 0] km s$^{-1}$, respectively. The white contours showing the 1.3\,mm continuum emission are [9, 13, 17, 50]$\times\sigma_{\rm 1.3\,mm}$ to emphasize the streamers. The white dashed line represents the direction of $\rm H_2$ knots \mycite{10.1046/j.1365-8711.2003.06419.x,2024A&A...684A..86B} while the yellow dashed line shows the direction of $^{12}$CO outflow. The orange dashed lines represent the streamers in 1.3\,mm continuum. \textcolor{red}{The outermost contours of the red and blue lobes are also shown in {B}.}
}
\label{dcn_con_streamer} 
\end{figure*}

\begin{figure*}[h]
\centering
\includegraphics[width=1\linewidth]{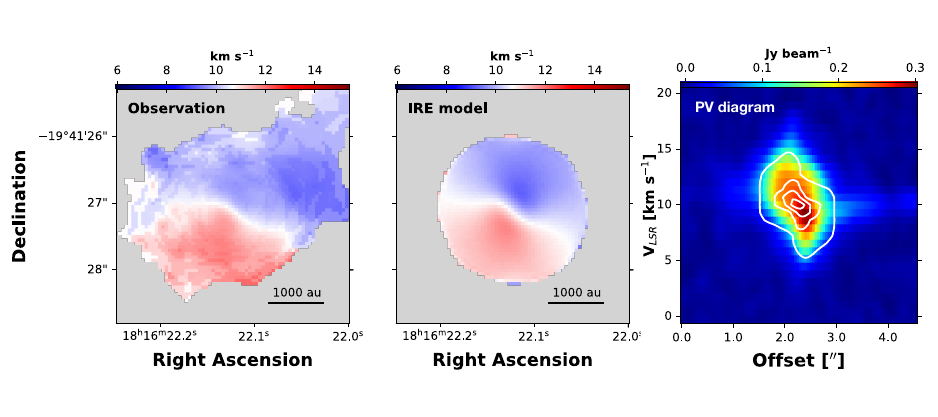}
\caption{ {The comparison between the kinematics of the CH$_3$OH observation and the best-fit model.}
The {left} and {middle panels} show the centroid velocity maps of the observation and the model, respectively. The PV diagrams of the observation (colormap) and the model (white contours) are shown in the {right panel}. Contour levels are 20\%, 40\%, 60\%, and 80\% of the peak intensity of the model.
}
\label{com_obs_model} 
\end{figure*}

\begin{figure*}[h]
\centering
\includegraphics[width=1.\linewidth]{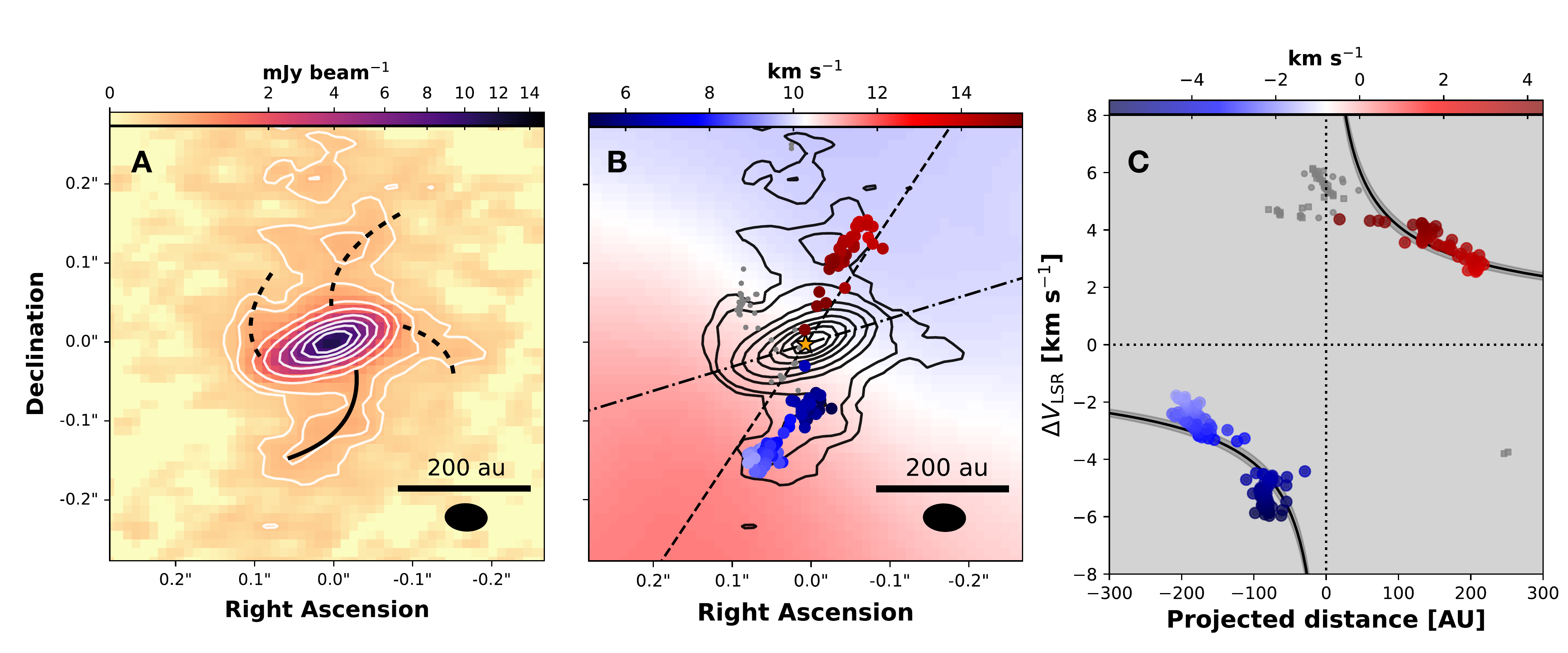}
\caption{
{ALMA long-baseline (1.3\,mm), QUARKS (1.3\,mm), and VLA 12\,GHz methanol maser observations.} {A,} ALMA archival long-baseline 1.3\,mm continuum observation ($\theta\approx30\,{\rm mas}$). The contour levels are at [3, 5, 15, 30, 60, 90, 120, 150]$\times\sigma_{\rm lb}$ where $\sigma_{\rm lb}$ is 0.06\,mJy per beam. The black curve sketches the southern mini-arm and the three dashed curves mark the tentative mini-arms in the disk. {B,} The intensity-weighted velocity map of $\rm CH_3OH~4_2-3_1~E1~vt=0$ (background) and 12 GHz methanol maser spots. The line-of-sight velocity of the selected maser spots is color-coded. The black dashed line indicates the direction of the PV cut for the maser spots which is also the direction of the largest velocity gradient of $\rm CH_3OH$ thermal line emission. The black dash-dotted line is the position angle of the inner disk from the 2-dimensional Gaussian fitting. The orange star marker represents the peak of the long-baseline continuum. The black contours are the same as in panel {A}. {C,} The PV diagram of 12\,GHz methanol maser projected to the PV cut in panel {B}. The black curves represent the Keplerian rotation curve with a scaled enclosed mass $M_{*,d}\times\sin^2i$ of 1.93\,$M_{\odot}$. The grey spots in panels {B} and {C} are the excluded maser spots. 
}
\label{maserdisk} 
\end{figure*}

\begin{figure*}[h]
\centering
\includegraphics[width=0.8\linewidth]{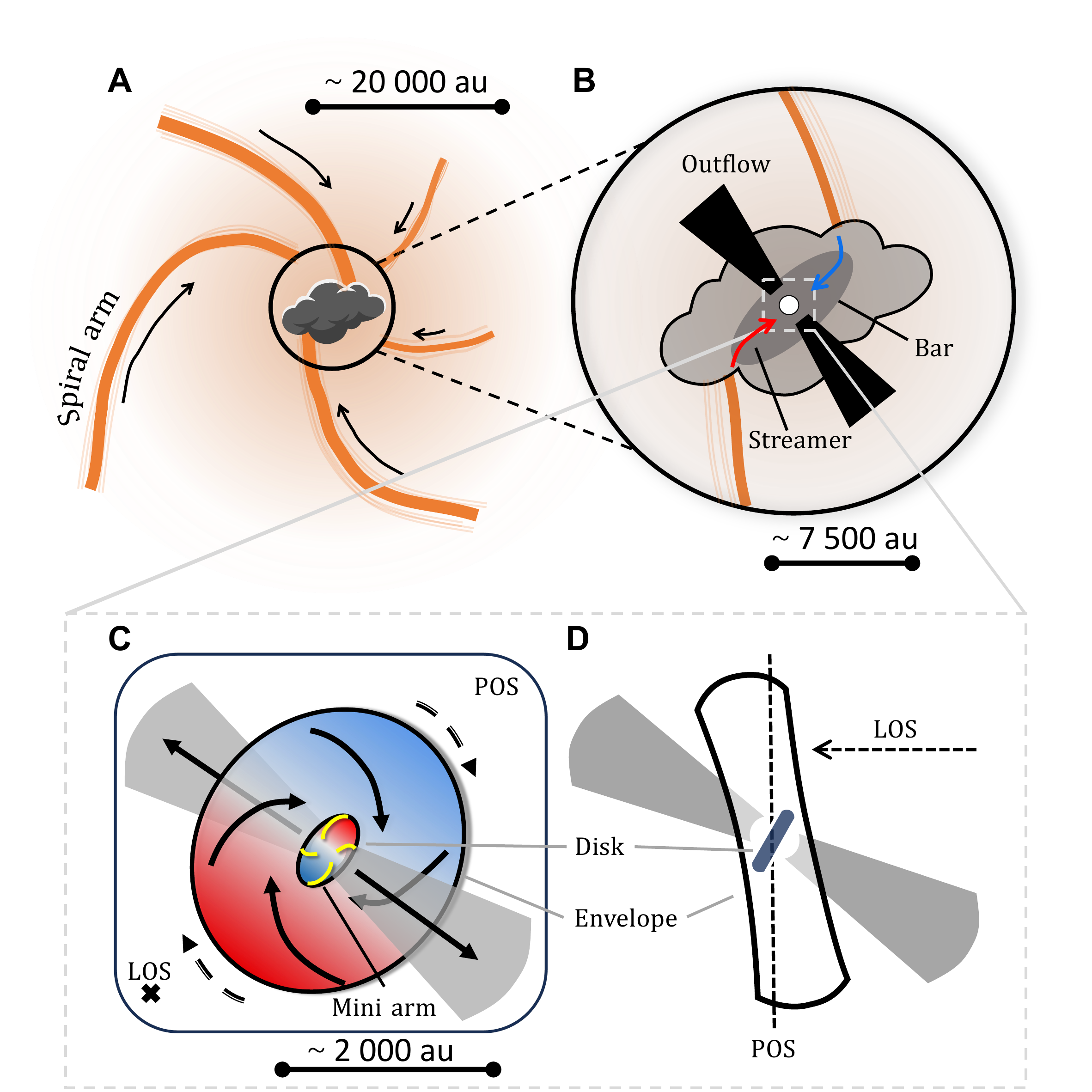}
\caption{{A schematic view of the complex system in IRAS 18134-1942.} \textcolor{red}{Panel {A}: Spiral arms with sizes of $\times$10,000 au. Panel {B}: The ``bar" ($\sim$7500 au) and streamers (blue and red arrows) revealed by 1.3 mm continuum emission. Black cones show molecular outflow traced by $^{12}$CO. Panel {C} and {D} show the front view and side view of the envelope as well as the embedded disk, respectively. `POS' denotes `Plane of Sky' while `LOS' means `Line of Sight'. The most remarkable feature is that the velocity distribution of the disk (inner ellipse; $\sim$100-400 au) traced by 12 GHz Class II methanol masers is opposite to the velocity field of the envelope/pseudodisk (outer ellipse; $\sim$1000-2000 au). The yellow curves in panel {C} show the mini-arms of the disk.} }
\label{cartoon}
\end{figure*}

At a distance of 1.25 kpc from the Sun \mycite{2014A&A...566A..17W}, MDC IRAS 18134-1942 (G11.497-1.485, hereafter I18134) is a massive cloud core containing a UC H{\sc ii} region \mycite{2003MNRAS.341..270V}, with a gas mass of 238\,$M_{\odot}$ enclosed in a radius of $\sim0.23$ pc \mycite{2018MNRAS.473.1059U}. This MDC hosts strong, highly variable Class II methanol \mycite{1993MNRAS.261..783S, 2005A&A...432..737P, 2010A&A...517A..56F,2000A&A...360..311S, 2004AJ....127.3479G, 2024A&A...684A..86B} and water maser emission \mycite{1991A&A...246..249P, 2007PASJ...59.1185S, 1996A&A...311..971C, 2010MNRAS.406.1487B}. In particular, an intense flare of the 6.7\,GHz methanol maser was detected recently, increasing by more than a factor of 10 over a four month period, indicating active high-mass star formation associated with possible intense accretion bursts \mycite{2024A&A...684A..86B}. 
We observed the source using the Atacama Large Millimeter/submillimeter Array (ALMA) at 3\,mm (Band 3, Fig.~\ref{cont_b3_b6}A) and 1.3\,mm (Band 6, Fig.~\ref{cont_b3_b6}B and \ref{dcn_con_streamer}) wavelengths. 
At resolutions of 0.03$\arcsec$--2$\arcsec$, corresponding to 37--2400\,au at the source distance, we discover a remarkable ``spirals-bar-disk-outflow'' complex in this MDC.
Thermal molecular line emission in conjunction with the Class II methanol masers reveals a pseudodisk (or a rotating and infalling envelope, Fig.~\ref{com_obs_model}) and a Keplerian disk surrounding the embedded HMYSO (Fig.~\ref{maserdisk}), which drives a bipolar outflow. The pseudodisk and the disk are fed by gas streamers that are embedded in a $\sim$7,500 au long bar-like structure (``bar"). The ``bar" is connected to three prominent $\sim$20,000 au long spiral arms as revealed by the 3 mm continuum emission. This discovery provides unprecedented insights into the complex interplay between the accretion disk and the hierarchical structures beyond it (Fig. \ref{cartoon}).\\

\noindent{\large \textbf{Results}}\\

\noindent{\textbf{A Spiral-like system at core scale}}

\begin{figure*}[h]
\centering
\includegraphics[width=0.8\linewidth]{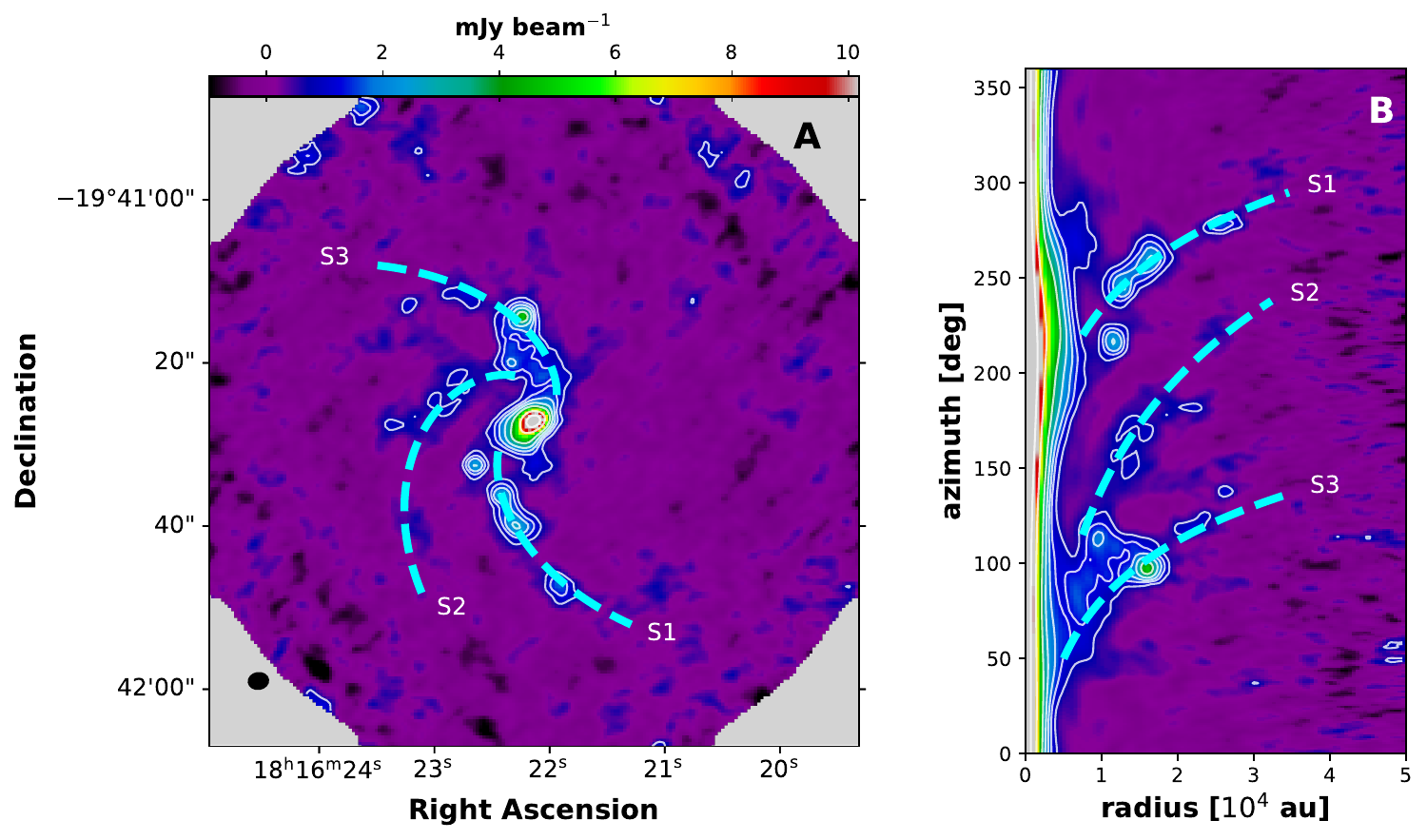}
\caption{{3\,mm continuum emission and the identified spiral arms.} {Panel A} shows 3\, mm continuum emission. {Panel B} shows the same emission as {panel A} but is projected with respect to the clump center and replotted as a function of radius and polar angle. The white contours at both panels are at the same levels as in Fig.~\ref{cont_b3_b6}.}
\label{b3cont_spiral}
\end{figure*}

\begin{figure*}[h]
\centering
\includegraphics[width=0.99\linewidth]{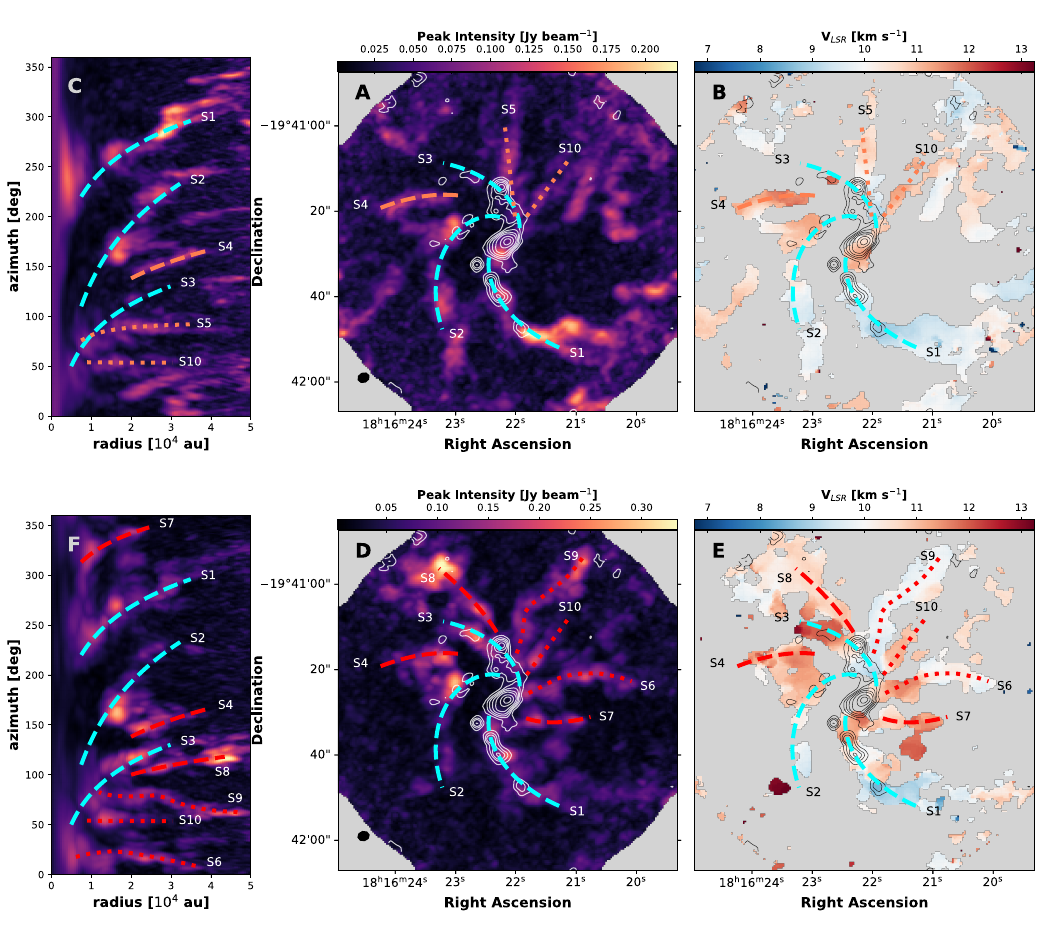}
\caption{\textbf{The spiral-like system traced by $\rm H^{13}CO^+$ and CCH.} {A,} The peak intensity map (8$^{\rm th}$ moment map)  of $\rm H^{13}CO^+$. The cyan dashed curves trace the spiral arms identified in the 3\,mm continuum emission. The spiral arms identified in $\rm H^{13}CO^+$ are plotted as orange dashed curves and the filaments outlined manually are plotted as coral dotted curves. {B,} The centroid velocity map, depicting the kinematics of the gas. {C} The polar projection of the peak intensity map {A}. {D-F} Same as {A-C}, but for CCH. The spiral arms identified in $\rm H^{13}CO^+$ are plotted as red dashed curves and the filaments outlined manually are plotted as red dotted curves. All the contours are 3\,mm continuum emission at the same levels as Fig.~\ref{cont_b3_b6}.}
\label{spiral_h13cop}
\end{figure*}

ALMA 3\,mm observations of continuum emission (Fig.~\ref{cont_b3_b6}A and  Fig.~\ref{b3cont_spiral}) and molecular lines of $\rm H^{13}CO^+$ and CCH (Fig.~\ref{spiral_h13cop}) reveal a distinct spiral system in I18134 that appears to channel material from beyond 20,000 au down to the central gravitational potential well. 
While spiral-arm-like structures induced by gravitational instability or by external perturbation such as flyby encounters are often witnessed in protostellar disks \mycite{2020NatAs...4..142L,2020NatAs...4.1170C,2022NatAs...6..837L,2023NatAs...7..557B,2024Natur.633...58S}, well-organized spiral arms in MDCs are rarely seen, with only a few candidates reported \mycite{2015ApJ...804...37L,2021ApJ...915L..10S,2022ApJ...941...51H,2023MNRAS.520.3259X,2023ApJ...959L..31O,2024MNRAS.530.1956S}. 

Three prominent dusty spiral arms are visible in 3\,mm continuum emission and can be modeled as logarithmic spirals ({Materials \& Methods}; see Fig.~\ref{b3cont_spiral}), which are further confirmed in molecular line emission (see  Fig.~\ref{spiral_h13cop}). One (S4) and two (S7-8) additional spiral arms can be identified in $\rm H^{13}CO^+$ and CCH, respectively, trailing in the same direction as the dusty arms (see  Fig.~\ref{spiral_h13cop}). Besides, four more spiral-like structures (S5, S6, S9, S10) are recognizable visually in $\rm H^{13}CO^+$ and CCH, deviating from the trailing pattern of the dusty spirals (see  Fig.~\ref{spiral_h13cop}). Different molecules might trace different parts of the cloud. To visualize the overall gas distribution, we stack the continuum emission and the peak intensity maps of $\rm H^{13}CO^+$ and CCH with uniform weighting, after normalizing each to their maximum (fig.~S\ref{spiral_stack}). The resulting composite map reveals an intricate spiral-like system.

\begin{figure*}[h]
\centering
\includegraphics[width=0.99\linewidth]{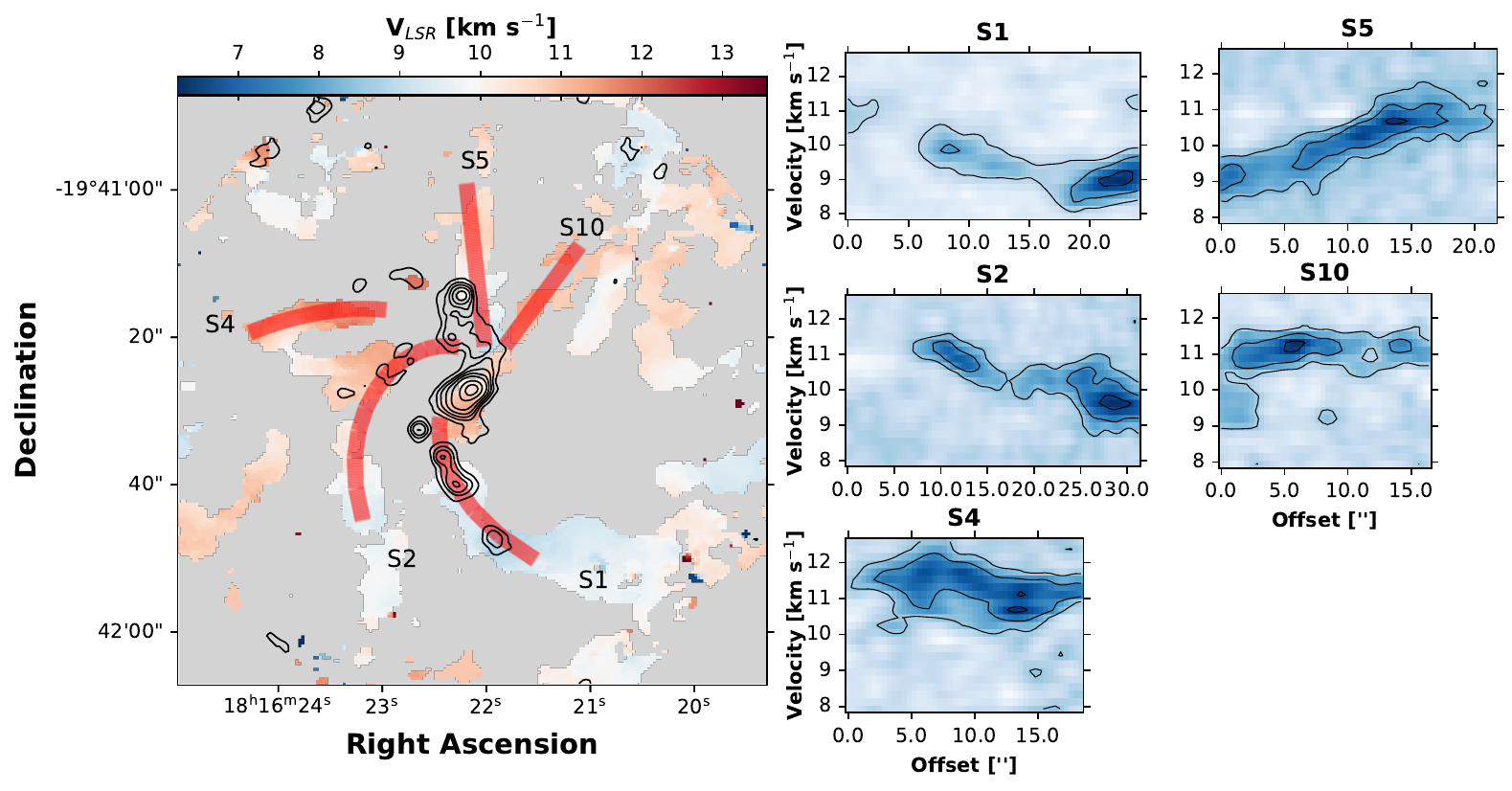}
\caption{{The velocity coherent filaments traced by $\rm H^{13}CO^+$.} {Left} shows centroid velocity map of $\rm H^{13}CO^+$, with 3\,mm continuum emission overploted. The filaments identified previously are shown as red bands. {Right} shows position-velocity diagrams along each streamers in $\rm H^{13}CO^+$. The offset is from the center outward. The contours are drawn at the level of 20$\%$, 50$\%$, and 90$\%$ of the peak intensity of the PV diagrams.}
\label{h13cop_pvs}
\end{figure*}

\begin{figure*}[h]
\centering
\includegraphics[width=0.99\linewidth]{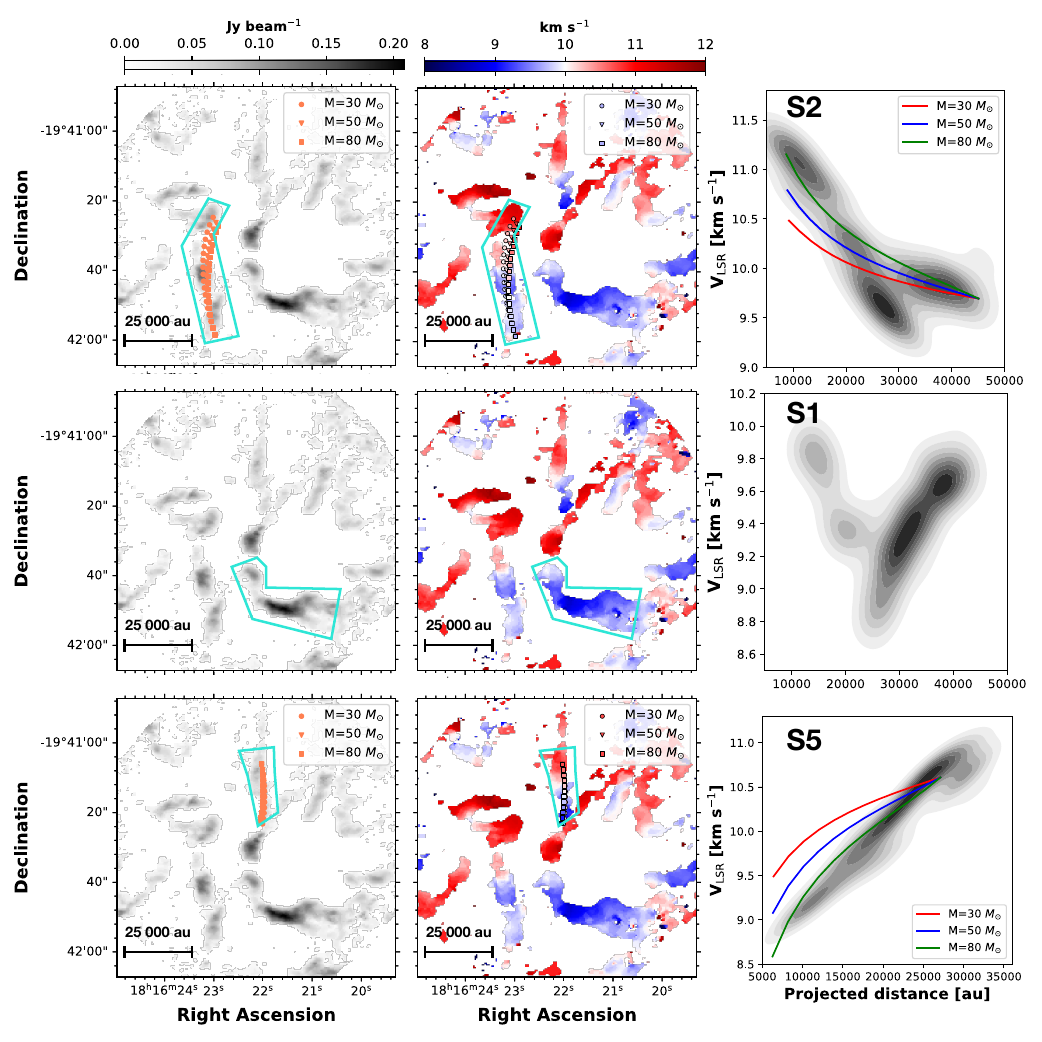}
\caption{{The kinematics observed in $\rm H^{13}CO^+$ and the model trajectories of the infall-rotating particle.} Rows correspond to S2 (first row), S1 (second row), and S5 (third row), while columns show peak intensity maps (first), centroid velocity maps (second), and kernel density distribution of centroid velocity versus projected distance (third).}
\label{infallrotate_h13cop}
\end{figure*}

The apparent spatial association alone is insufficient to conclusively validate these as bona fide structures; coherent kinematics are also needed. The intensity-weighted velocity maps of $\rm H^{13}CO^+$ and CCH (see right panels in  Fig.~\ref{spiral_h13cop}) and the position-velocity (PV) diagrams (see  Fig.~\ref{h13cop_pvs} and fig.~S\ref{cch_pvs}) along the spiral arms present smooth velocity gradients or flat velocity profiles along the identified narrow structures, confirming their velocity-coherent nature. We further model the spiral arms in $\rm H^{13}CO^+$ as the streamlines of a particle with an initial rotation, subject to the gravity of a central massive object \mycite{2009MNRAS.393..579M, 2019NatAs...3..517Z, 2020NatAs...4.1158P} (Fig.~\ref{infallrotate_h13cop}). 
While this idealized kinematic model with a central mass of 50-80 $M_{\odot}$ successfully reproduces the gas kinematics along S2 and S5, it can not capture the entire system's dynamics, particularly the V-shaped velocity profiles of S1 and S2 that are likely indicative of the gravity-driven accretion of the cores lying within the filaments. 
Nevertheless, the velocity coherence along the spiral arms, coupled with the partial agreement between the model and the kinematics supports the idea that the gas flows in along the spiral arms with a mass inflow rate of $\sim3.4\times10^{-4}$ M$_\odot$ yr$^{-1}$ ({Materials \& Methods}). This mass inflow rate is comparable with other high-mass star-forming regions with similar scale reported in the literature \mycite{2018ApJ...855....9L, 2021ApJ...915L..10S,2022ApJ...936..169R, 2024MNRAS.533.1075Z}\\

\vspace{0.5em}

\noindent{\textbf{A bar-like structure}}

A prolate structure (the central ellipse in Fig.~\ref{cont_b3_b6}A) with a aspect ratio $\sim$1.5 ($\sim$ 5300$\times$3500\,au) has formed in the center of I18134, fed by the 20,000-au-scale spiral-like system. The prolateness of the structure is more distinctly visible in the composite map (fig.~S\ref{spiral_stack}). It is further resolved in 1.3\,mm continuum emission (Fig.~\ref{cont_b3_b6}B) and molecular line emission (e.g., DCN, Fig.~\ref{dcn_con_streamer}B) at a high resolution of 0.3$\arcsec$, showing a more prominent bar-like elongated structure (hereafter denoted as the ``bar"). 
The ``bar" is resolved into a bright compact object (called the ``nucleus") with slender extensions, referred to as streamers, stretching toward the southeast and northwest. The slender streamers and the nucleus resemble a bar-like envelope linking the spiral system to the central protostar. 

The spatial distribution of DCN strongly correlates with the 1.3\,mm continuum emission (Fig.~\ref{dcn_con_streamer}), tracing a similar bar-like structure. The centroid velocity map of DCN (Fig.~\ref{dcn_con_streamer}B) shows that the northern spiral arm is blueshifted relative to the systematic velocity of I18134, suggesting a gas streamer infalling from the north to the ``bar". Additionally, the gas kinematics
along the ``bar" exhibit red-blueshifted symmetry around the ``nucleus", further supporting the idea that these streamers are transporting gas and converging toward the ``nucleus''. This hypothesis is strengthened by the PV diagrams of various molecules cutting along the ``bar" (fig.~S\ref{pvs}). Among these, DCN, H$_2$CO, HC$_3$N, and C$^{18}$O trace an elongated velocity-coherent structure spanning $\sim6\arcsec$ (fig.~S\ref{pvs} and~S\ref{intensity_profile}), indicating that the ``bar" may extend to a distance of $\sim$7,500 au. The mass inflow rate from streamers to the ``nucleus" is estimated to be $\sim2.6\times10^{-4}$ M$_\odot$ yr$^{-1}$ ({Materials \& Methods}), which is similar to the mass inflow rate along the spiral arms. Therefore, the ``bar" plays a key intermediary role in channeling gas from the core-scale spiral-like system and fueling the very central ``nucleus".


\vspace{0.5em}

\noindent{\textbf{Streamers penetrate the ``bar" and connect to the nucleus}}

Zooming further into the ``bar", both DCN and CH$_3$OH show asymmetric spatial distribution (fig.~S\ref{dcn_ch3oh}). The distribution of DCN consists of three components: the diffuse ambient gas that forms the main body of the ``bar", a compact point source associated with the ``nucleus", and a small tail connecting to the ``nucleus". Notably, this tail extends along the southeast streamer. At the same position, a less apparent tail is also visible in the CH$_3$OH emission. Additionally, a second tail appears west of the ``nucleus", extending along the northwest streamer. The alignment of the tails and dusty streamers may suggest that the tails represent the terminal ends of the streamers where the gas is transported deep into the ``nucleus". 

All the PV diagrams along the ``bar" (fig.~S\ref{pvs}), except for that of C$^{18}$O, exhibit a diamond-shaped feature peaked at the ``nucleus".
These diamond-shaped PV diagrams are characteristic of an infalling-rotating envelope (IRE), e.g. a non-rotationally supported envelope/pseudodisk \mycite{2014ApJ...791L..38S, 2015ApJ...812...59O,2016ApJ...824...88O, 2017ApJ...837..174O,2019ApJ...873...73Z}. The continuity from the streamers to the pseudodisk suggests that the streamers are transporting material to a pseudodisk and the gas starts to rotate alongside the pseudodisk. We model the pseudodisk traced by CH$_3$OH thermal emission as an IRE to infer the geometry of the pseudodisk and the mass of the central mass (Fig. \ref{com_obs_model} and  fig.~S\ref{chi2}). Despite the large uncertainty of the central mass, the model indicates a nearly face-on pseudodisk with an inclination $<15^{\circ}$. Similar geometry and kinematics of the pseudodisk are also well revealed by other dense gas tracers such as CH$_3$CN lines (fig.~S\ref{ch3cn_mom_pv}).

\vspace{1em}

\noindent{\textbf{The misaligned disk}}


Zooming into the ``nucleus'' with ALMA long-baseline observation at 1.3\,mm  (angular resolution of $\theta\approx0.03''$, equivalent to 37\,au), one single compact source which could be the disk is detected inside in the center of the ``nucleus'' (Fig.~\ref{maserdisk}A). Besides the central compact component, one clear ``mini-arm'' can be identified in the south, while another three are tentatively detected in the north, west, and east, similar to the four-arm disk reported in \mycite{2023NatAs...7..557B}. On the basis of 2-dimensional Gaussian fitting to the disk, the size and inclination of the disk are $\sim$110\,au and $\sim75^{\circ}$, respectively. This inclination differs from the pseudodisk by more than 60$^{\circ}$. Additionally, the presence of $^{12}$CO bipolar outflow suggests the disk is not face-on. The combined evidence indicates that the disk and the pseudodisk are notably misaligned.

Jansky Very Large Array observations with a resolution of $\sim0.2\arcsec$ led to the detection of 12 GHz Class II methanol masers oriented along the southeast-to-northwest direction \mycite{2024A&A...684A..86B}, perpendicular to the CO bipolar outflow, indicating that the Class II methanol masers may trace the disk (Fig. \ref{maserdisk}b). 
Specifically, the blueshifted maser cluster corresponds to the southern spiral arm seen in the long-baseline observation, the redshifted cluster aligns with the northern arm, and another cluster (in gray color) seems to align with the outflow (Fig. \ref{maserdisk}b). 
The linear distribution of the Class II methanol masers suggests that the disk is likely viewed at a larger inclination than the pseudodisk. This further supports that the disk is misaligned with the pseudodisk. Meanwhile, the orientation of the linear distribution likely corresponds to the major axis of the disk traced by the maser which also deviates from the major axis of the continuum disk. This might suggest that the masers and the long-baseline continuum trace different disk structures. For example, the maser and the continuum trace the outer and inner disk, respectively, and they might also be misaligned.

\textcolor{red}{The most remarkable feature is that the velocity distribution of the masers is opposite to the velocity field of the pseudodisk (see Fig.~\ref{maserdisk}C and the cartoon in Fig.~\ref{cartoon}C).} The velocity reversal may suggest opposite signs for the inclination of the disk traced by the maser and the pseudodisk. Previous observations have detected misalignment of the orientation between the disk and the envelope in low-mass star forming cores \mycite{2022ApJ...927...54O, 2007A&A...475..915B, 2007A&A...461.1037B}.  In extreme cases, the Keplerian disk could be counter-rotating relative to the pseudodisk. While mechanisms for this extreme scenario have been proposed in low-mass star-forming regions: the interaction with the spiral arm in a circum-multiple disk \mycite{2021PASJ...73L..25T} and non-ideal magneto-hydrodynamics effects \mycite{2011ApJ...738..180L, 2017PASJ...69...95T}, it is still theoretically unclear whether this scenario could happen in massive disks. Anyway, here we provide a firm detection of a misaligned disk surrounding a HMYSO. 


The opposite signs for the inclination could also arise from the nearly face-on orientation of the pseudodisk and the potential wobbling due to the non-uniform angular momentum transfer to the Keplerian disk via asymmetric accretion flows, leading to a few tens of degrees of shift in the spin axis orientation \mycite{2017ApJ...839...69M, 2024arXiv240506520P}. \textcolor{red}{To better visualize this effect, fig.~S\ref{orien_schem} provides a schematic representation of this inclination difference.} After excluding the maser cluster aligned with the outflow direction, the remaining 12 GHz Class II methanol masers follow Keplerian rotation very well ({Materials \& Methods} and Fig.~\ref{maserdisk}c). We perform Keplerian curve modeling on the spatial-kinematics of the masers in the projected position-velocity diagram.
The best-fit scaled enclosed mass ($M_{*,d}\times\sin^2 i$ where $M_{*,d}$ is the dynamic mass and $i$ is the inclination of the disk) is 1.93~$M_{\odot}$. If we take the best-fit inclination of the pseudodisk as the inclination of the disk, the mass of HMYSO $M_{*,d}$ would be over 180~$M_{\odot}$ which is almost the total mass of the MDC. Thus, the inclination of the maser disk should be larger than that of the pseudodisk, which further suggests the misalignment between the pseudodisk and the maser disk. 
Radio continuum emission from the central UC H{\sc ii} region suggests that the spectral type of the central star is between B3 and B0 \mycite{2003MNRAS.341..270V}, corresponding to a stellar mass of 8-16\,$M_{\odot}$ \mycite{2013ApJS..208....9P}. Assuming this stellar mass range, the inclination of the maser disk would be $20^{\circ}$ to $30^{\circ}$.

As mentioned above, the ``nucleus'' in MDC IRAS 18134-1942 is fed by gas streamers within the elongated bar-like structure, indicating that the formation of the misaligned disk may be induced by gas infall, as suggested in numerical simulations \mycite{2021A&A...656A.161K}. Multiple or episodic gas acquisition events through the large-scale barred spiral-like system during the gas accretion and angular momentum build-up of the disk may naturally lead to the kinematical misalignment (or even counter-rotation) of components in the “nucleus”.

In addition, the shock tracers, $\rm H_2$, SiO thermal emission, and 22 GHz water masers, are well-aligned, but they deviate by approximately $30^{\circ}$ from the direction of the $^{12}$CO outflow (left panel in Fig.~\ref{dcn_con_streamer}C and  fig.~S\ref{outflow_sio_h2_maser}).
The wobbling disk due to the asymmetric accretion flow could also be the reason for the misalignment between the $^{12}$CO outflow and the shock, and the potential jet precession revealed by $^{12}$CO high-velocity emission (fig.~S\ref{co_channel}).

The disk-mediated accretion rate of the HMYSO, inferred from CO outflow emission, is $\sim$2.8$\times10^{-6}\,M_{\odot}~{\rm yr}^{-1}$({Materials \& Methods}), which is two orders of magnitude smaller than the mass inflow rate along the spiral arms and the streamers. Since we assume $^{12}$CO emission is optically thin, and due to the potential missing flux caused by the lack of short spacing of the interferometer, this accretion rate would be the lower limit of the exact accretion rate. However, the estimated accretion rate is consistent with the accretion rates of high-mass star formation regions reported in \mycite{2020ApJ...903..119L}. In particular, a strong correlation between the accretion rate $\dot{M}_{\rm acc}$ and luminosity-to-mass ratio $L_{\rm bol}/M_{\rm clump}$ is reported: $\log(\dot{M}_{acc})=0.8\log(L_{\rm bol}/M_{\rm clump})-5.8$. With a $\log(L_{\rm bol}/M_{\rm clump})$ of 0.65 \mycite{2018MNRAS.473.1059U}, $\log(\dot{M}_{\rm acc})$ of I18134 from this relation would be -5.28 which is closed to the estimated $\log(\dot{M}_{\rm acc})$ of -5.55. The drastic decrease from the mass inflow rate of the streamers to the disk-mediated accretion rate suggests that the rotating envelope and protostellar disk may play a crucial regulatory role in the mass growth of the protostar.

\vspace{1em}
\noindent{\large \textbf{Discussion}}

The discovery of the ``spirals-bar-disk-outflow'' complex (see the cartoon in Figure \ref{cartoon}) in MDC IRAS 18134-1942 provides the clearest observational evidence that HMYSOs can efficiently assemble their masses through multi-scale gas accretion within a dense core ($\sim$0.1 pc). Gas flows in from spiral arms down to the ``bar", and then finally onto the disk. 
The gas transfer rate remains relatively constant ($\sim 10^{-4}\,M_{\odot}~{\rm yr}^{-1}$) from the spiral system to the ``bar'', suggesting a hierarchical but continuous mass infall process.
The asymmetric gas inflows, ending at the disk, could trigger the accretion bursts of the HMYSOs \mycite{2023ApJ...954L..25B,2024ApJ...966..119L}, which in turn could explain the methanol maser flares. Such inflows could also inject angular momentum non-uniformly to the HMYSO, resulting in the disk wobbling and the jet precession. The disk wobbling could cause the misalignment of the disk and envelope. 
Our discovery manifests that barred spiral-like structures within MDCs play an important role during the gas accretion and angular momentum build-up of massive disks. Future studies of more similar systems will be crucial for unraveling the mystery of high-mass star formation as well as massive disk formation. Furthermore, from the visual impression alone, the ``spirals-bar-disk-outflow" complex (Figure \ref{cartoon}) closely resembles the morphology of a barred spiral galaxy with an active galactic nucleus (AGN) despite their very different physical scales and central engines (HMYSO versus supermassive black hole). \textcolor{red}{This work demonstrates that spiral systems and bar-like structures can be formed not only in self-gravitating rotating systems like galactic discs (for example, see some state-of-the-art simulation of the formation of galactic discs in very gas-rich environments \mycite{2024MNRAS.535..187T, 2024ApJ...968...86B, 2025arXiv250201895B}) but also in high-mass star formation regions. This suggests that in certain systems these structures are closely linked to the mass transport process}. Detailed comparison to barred spiral galaxies is beyond the scope of this work. However, insight from the formation and evolution of galaxies and AGNs would provide valuable inspiration for the high-mass star formation community, and vice versa.
%

\vspace{1em}


\clearpage
\phantomsection
\label{sec:methods}
\noindent{\large \textbf{Materials \& Methods}}


\renewcommand{\figurename}{Fig.}
\renewcommand{\figureautorefname}{Fig.}
\renewcommand{\tablename}{Table.}
\renewcommand{\tableautorefname}{Table.}


\vspace{0.5em}
\noindent\textbf{\large ALMA observations and data reduction}\\

\noindent\textbf{ALMA 3\,mm observations}~~

\target was observed under the project ``ALMA Three-millimeter Observations of Massive Star-forming regions (ATOMS)" with Atacama Compact 7\,m Array (ACA) and 12\,m ALMA arrays (C43-3 configuration) using the Band\,3 receiver \mycite{2008SPIE.7020E..1BC} on 2019 November 1-3 (Project ID: 2019.1.00685.S; PI: Tie Liu). 
The standard calibration was done using Common Astronomy Software Applications (CASA) package \mycite{2007ASPC..376..127M} 5.6. 
\textcolor{red}{The ACA data and 12\,m-array data were combined and imaged using the same version of CASA with natural weighting, yielding an angular resolution of 2.15 arcsec $\times$ 1.72 arcsec (position angle of -81.4$^{\circ}$ ) for the 3\,mm continuum emission.}
The continuum image is shown in Fig.~\ref{cont_b3_b6}; \textcolor{red}{The pixel size is 0.4$^{\prime\prime}$} and it achieved an rms noise of $\rm\sim0.25\,mJy\,beam^{-1}$. We refer to the survey description paper \mycite{2020MNRAS.496.2790L} for the observations and data reduction details.\\

\noindent\textbf{ALMA 1.3\,mm observations}~~

We conducted ALMA Band\,6 \mycite{2004stt..conf..181E} observations of \target ACA and 12\,m arrays (C-2 and C-5 configurations) under the project ``Querying Underlying mechanisms of massive star formation with ALMA-Resolved gas Kinematics and Structures (QUARKS)" (project code: 2021.1.00095S, PIs: Lei Zhu, Tie Liu, and Guido Garay) on 2023 April 17 (ACA), 2024 March 19 (C-2) and 2022 August 15 (C-5). The calibration, image, and self-calibration were done using CASA 6.5. \textcolor{red}{We cleaned the continuum image and the cube data with a pixel size of 0.05$^{\prime\prime}$ using the \textit{briggs} weighting. The \textit{robust} parameter is adopted as 0.5. We conducted three rounds of phase-only self-calibration with a shortest solution interval of 12 s}. For the details of observations and data reduction, we also refer to the survey description paper \mycite{2024RAA....24b5009L}. We achieved an angular resolution of 0.30 arcsec $\times$ 0.27 arcsec (position angle of 66.8$^{\circ}$) and an rms noise on the continuum emission of $\rm\sim0.25\,mJy\, beam^{-1}$. For the highest-resolution observation, we use the self-calibrated product image from the ALMA science archive (project code: 2021.1.00455.T, PI: Todd Hunter). The synthesized beam size is $\rm 0.05\,arcsec\times0.03\,arcsec$ (position angle of 87.4$^{\circ}$). The rms noise is $\rm \sim0.06\,mJy~beam^{-1}$.\\

\vspace{0.5em}
\noindent\textbf{\large Spiral-arm-like structures}\\

\noindent\textbf{Morphology}
\label{sec:streamer_morphology}

We first characterize spiral-like filaments (hereafter spiral arms) on the 3\,mm continuum map. For simplicity, we model the spiral arms as logarithmic spirals, described by $R(\theta)=R_0e^{a\theta}$, where $R_0$ is the starting radius of the spiral, $a$ is the growth rate of the spiral, and the pitch angle is given by $\arctan{|a|}$. The starting radius $R_0$ of each spiral arm is manually determined from the projected polar plot. The polar plots were made via Python package \textit{polarTransform}. Using the logarithmic spiral curve, we generated spiral model images via the Bresenham Line Algorithm and convolved them with a 2-dimensional Gaussian kernel with the same full width at half maximum (FWHM) as the synthesized beam (fig.~S\ref{spiral_model_continuum}A). By calculating the Pearson correlation coefficient between the model images and the 3\,mm continuum map, we identified the spiral model that best matches the original continuum map (fig.~S\ref{spiral_model_continuum}B). The best fit of $a$ is robust against the choice of the FWHM. The best match spiral arms S1, S2, and S3 are outlined as cyan dashed curves in Fig.~\ref{cont_b3_b6}. The pitch angles of S1-3 are 49$^{\circ}$, 34$^{\circ}$, and 52$^{\circ}$, respectively. The pitch angles of S1 and S3 are quite similar which might suggest that the two arms form a symmetric grand-design spiral system.

The molecular gas distribution also exhibits a similar spiral pattern.  Fig.~\ref{spiral_h13cop} presents the peak intensity and centroid velocity maps of $\rm H^{13}CO^+$ $J=1-0$ and CCH $N_{J,K}=1_{3/2,2}-0_{1/2,1}$. The spiral pattern of the gas distribution is more distinctly visible in molecular gas. We employ a similar approach to identify and characterize the spiral arms in the peak intensity maps. The emission from S2 is more extended in $\rm H^{13}CO^+$ than in the continuum emission. The parameter $a$ of S2 measured in the continuum map is replaced by the best match $a$ when matching the model with the emission from $\rm H^{13}CO^+$, assuming that both map the same gas. One (S4) and three (S4, S7, S8) additional spiral arms are identified in $\rm H^{13}CO^+$ and CCH, respectively, as indicated by the orange and red dashed curves. While a logarithmic spiral curve was used to depict the gas distribution for simplicity, it can not fully capture the entire morphology. We outline manually additional filamentary structures in these peak intensity maps, as orange and red dotted lines. To examine the overall gas distribution, we stack the normalized peak intensity maps of $\rm H^{13}CO^+$ and CCH alongside with the 3\,mm continuum map with uniform weighting in fig.~S\ref{spiral_stack} (this neglects the relative abundance between dust, $\rm H^{13}CO^+$, and CCH).\\

\clearpage

\noindent\textbf{ Gas Kinematics}

The PV cuts are placed along the curves we identified in the previous section. The PV diagrams are shown in  Fig.~\ref{h13cop_pvs} and fig.~\ref{cch_pvs}, which reveal that the identified structures are velocity-coherent filaments. The velocity gradients seen in the PV diagrams may indicate that the gas is infalling to the center of the clump driven by the gravitational force of the central object.

\clearpage

\vspace{0.5em}
\noindent\textbf{Infalling-rotating particle trajectory}
\label{sec:particletrajetoty}

We model the kinematics of the filaments observed with $\rm H^{13}CO^+$ to validate whether the velocity gradient is consistent with infall motion using an analytic solutions \mycite{2009MNRAS.393..579M, 2020NatAs...4.1158P} of a particle moving within a rotating cloud toward a central mass. The model provides both the plane-of-sky position and the line-of-sight velocity along an infalling streamline. The model is defined by a central point mass, the inclination angle of the angular momentum vector of the rotating plane, and the angular velocity of the cloud assuming a rigid body rotation. The major inputs are the initial position and velocity of the particle in spherical coordinates.

We manually adjust the model and streamline parameters to find the best match parameter set. Based on the observed gas morphology, we assume the inclination angle to be between 0$^{\circ}$ to 20$^{\circ}$ (where 0$^{\circ}$ represents a face-on view). 
\textcolor{red}{The model that best reproduces the observed features is obtained using} the initial distance of 52,000 au, the position angle of 21$^{\circ}$, and the polar angle of 70$^{\circ}$ for S2, and the initial radius of 35,500 au, the \textcolor{red}{azimuthal} angle of 176$^{\circ}$, and the polar angle of 40$^{\circ}$ for S5. The position angle is measured from the north in a clockwise direction, and the polar angle is measured from the spin axis of the plane whose inclination angle is 10$^{\circ}$. The angular velocity is assumed as $4\times10^{-14}~{\rm rad~s^{-1}}$ \textcolor{red}{and the initial radial velocity is 0.6\,$\rm km~s^{-1}$. }
 Fig.~\ref{infallrotate_h13cop} shows the streamline trajectories (in coral) assuming a central mass of 50-80$\,M_{\odot}$ in S2 and S5 which spatially align with the emission in the peak intensity map (left column). The middle column shows the centroid velocity and also the trajectories color-coded by line-of-sight velocity. The right column presents the kernel density estimate (KDE) of the centroid velocity versus the projected distance from the central mass measured in the middle column. 

\clearpage

\vspace{0.5em}
\noindent\textbf{\large IRE modeling}\\

Based on the spatial distribution and velocity map, the $\rm CH_3OH~4_2-3_1~E1~vt=0$ emission is expected to trace the pseudodisk.
Here we model the pseudodisk as a flattened infalling and rotating envelope (IRE) with the publicly available code FERIA \mycite{2022PASP..134i4301O}. The central mass $M_{*,e}$, the centrifugal radius $r_{\rm CB}$, the outer radius of the IRE $r_{\rm out}$ and the inclination $i$ are treated as free parameters. The position angle of the IRE model is fixed to the velocity gradient observed in the centroid velocity map. Note that the uncertainties arising from the simplicity of the model could dominate over the observational noise. For this reason, instead of conducting a precise fitting method like MCMC, we employed a $\chi^2$ grid search to explore the parameter space in this study. To reduce computational time, we began with a coarse $\chi^2$ grid search to constrain the parameter ranges, followed by a fine grid search. The parameter ranges for the fine grid search are summarized in  table~S\ref{parameterrange}. \textcolor{red}{The best-fit parameter set with the lowest $\chi^2$ value} includes $M_{*,e}$ of 45\,$M_{\odot}$, $i$ of $6^\circ$, $r_{\rm CB}$  of 60\,au, and $r_{\rm out}$ of 950\,au (fig.~S\ref{chi2}). As shown in Fig.~\ref{com_obs_model}, the best-fit model can reproduce the overall kinematics, especially the curvature shown on the centroid velocity maps--representing the local velocity$\sim 10.6$\,km s$^{-1}$--due to the infalling and rotating kinematics. Notably, the $\chi^2$ distribution in the $i$ -- $M_{*,e}$ space reveals a ridge-like pattern, where central mass and inclination are correlated. The model is insensitive to central mass but tightly constrains the inclination to $<15^\circ$. The model only considers a central point mass, and the best-fit 45\,$M_{\odot}$ is too massive for a protostar at a very early stage. Consequently, despite the uncertainty in the central mass, the IRE modeling robustly indicates a nearly face-on pseudodisk.


\vspace{0.5em}
\noindent\textbf{\large Dynamics of the maser disk}

We manually selected two blueshifted and redshifted maser clusters in the southeast and northwest for the reason that the northeast cluster (grey spots in Fig.~\ref{maserdisk}A) is possibly associated with the outflow due to the similar velocity and direction compared with the 22 GHz water masers and the $\rm H_2$ knots \mycite{2024A&A...684A..86B} (see  fig.~S\ref{outflow_sio_h2_maser}), or it is associated with the tentative eastern spiral arm. After projecting the maser spot onto a position-velocity cut, we modeled the spatial-kinematics of the maser using a Keplerian rotation model, treating the scaled enclosed mass ($M_{*,d}\times\sin i^2$ where $M_{*,d}$ is the dynamic mass and $i$ is the inclination of the disk), and the system velocity as free parameters. The position angle of the position-velocity cut was fixed as the direction of the steepest velocity gradient in the ALMA observation. The disk center was assumed to be the peak of the long-baseline 1.3\,mm continuum. The best-fit Keplerian rotation profile is shown in the position-velocity diagram in Fig.~\ref{maserdisk}. The scaled enclosed mass was estimated using the least-square fitting, implemented in SciPy, to be $M_{*,d}\times\sin^2 i=1.93\pm0.07\,M_{\odot}$. \\

\vspace{0.5em}
\noindent\textbf{\large Misalignment between outflow and shocked gas}

We detect the outflow emission traced by $^{12}$CO covered in our frequency range.  fig.~S\ref{outflow_sio_h2_maser}A shows the red- and blueshifted outflow gas integrated between [20, 50] km s$^{-1}$ and [-30, 0] km s$^{-1}$, respectively, where it is free of $^{13}$CO and C$^{18}$O emissions. Along with the $^{12}\rm CO$ outflow gas, the shock-induced SiO, $\rm H_2$, and 22\,GHz water maser emission are also shown in the same figure. SiO, $\rm H_2$, and 22\,GHz water masers share a similar extended direction $\sim 253^{\circ}$, implying a common shock orgion, likely driven by the jet. The $^{12}\rm CO$ outflow is in a different direction, differing from the jet direction by $\sim30^{\circ}$.\\

\clearpage

\vspace{0.5em}
\noindent\textbf{\large Mass assembly rate at different scales}
We determine the column density of $\rm H^{13}CO^+$ $J=1-0$, CCH $N_{J,K}=1_{3/2,2}-0_{1/2,1}$ assuming optically thin conditions with \mycite{2015PASP..127..266M}:

\begin{equation}
    N=\frac{3h}{8\pi^3|\mu_{lu}|^2}\frac{Q_{\rm rot}}{g_u}\exp\left(\frac{E_u}{kT_{\rm ex}}\right)\left[\exp\left(\frac{h\nu}{kT_{\rm ex}}\right)-1\right]^{-1}\frac{\int T_{\rm B} d\nu}{T_{\rm ex}},
\end{equation}
where $h$, $|\mu_{lu}|^2$, $Q_{\rm rot}$, $g_{\rm u}$, $E_{\rm u}$, $k$, $T_{\rm ex}$, $\nu$, and $T_{\rm B}$ are the Planck constant, the dipole matrix element, the rotational partition function, the level degeneracy, the upper state energy, the Boltzmann constant, the excitation temperature which is adopted as 20 K, the frequency of the transitions, and the brightness temperature. The $|\mu_{lu}|^2$, $Q_{\rm rot}$, $g_{\rm u}$, and $E_{\rm u}$ are taken from CDMS Database \mycite{2001A&A...370L..49M}. The mass of the inflow is then calculated through
\begin{equation}
    M_{\rm inflow}=X^{-1}m_{\rm H_2}AD^2\sum_{i,j}N_{ij},
\end{equation}
where $X$, $m_{\rm H_2}$, $A$, $D$, $N_{ij}$ are the molecular abundance relative to $\rm H_2$, the mass of a hydrogen molecule, the angular area of a pixel, the source distance, and the column density at the pixel$_{ij}$. We calculate the mass of the inflows seen in $\rm H^{13}CO^+$ emission and the rest seen in CCH emission. We assume the abundances of H$^{13}$CO$^+$ and CCH to be 1.28$\times10^{-10}$ \mycite{2013ApJ...777..157H} and 3.72$\times10^{-8}$ \mycite{2012ApJ...756...60S}, respectively.
The total masses of the H$^{13}$CO$^+$ and CCH spiral system are 22.9 and 3.4\,$M_{\odot}$, respectively. Assuming that the velocity gradient $\nabla{V}_{||}$, which is the velocity profile in the Moment 1 maps, is due to the inflow, we further calculate the inflow rate $\dot{M}_{||}$ via \mycite{2013ApJ...766..115K}:
\begin{equation}
    \dot{M}_{||}=\frac{\nabla{V}_{||}M_{\rm inflow}}{\tan{\theta}},
    \label{eq:3}
\end{equation}
where $\theta$ is the inclination angle. We assume the inclination angle to all be 45$^{\circ}$ except for S2 and S5, which are 30$^{\circ}$ and 60$^{\circ}$, respectively, based on the previous infall-rotating particle trajectory modeling.
The total inflow rate for the spirals detected in $\rm H^{13}CO^+$ and CCH is 3.4$\times10^{-4}\,M_{\odot}~{\rm yr}^{-1}$. 

\begin{figure*}[h]
\centering
\includegraphics[width=1\linewidth]{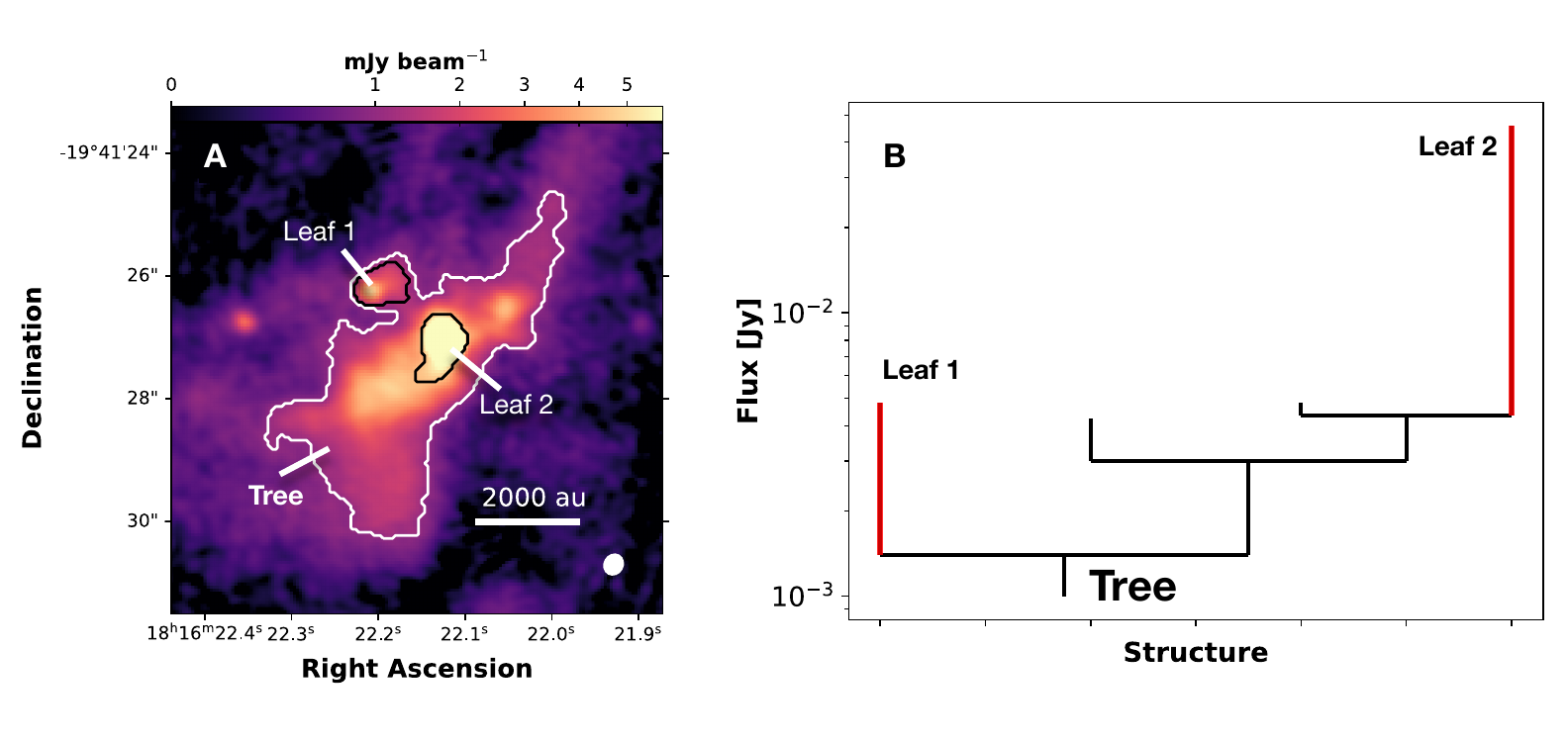}
\caption{ {The hierarchical structure of the ``bar''.}
{A,} The 1.3\,mm continuum overlaid with the ``tree'' structure as white contour and ``leaves'' structure as black contours. The most luminous leaves \textit{Leaf 1} and \textit{Leaf 2} are considered as protostars. {B,} The dendrogram of the ``bar'' identified by \textbf{astrodendro}. 
}
\label{cont_dendro}
\end{figure*}

We estimate the total mass of the ``bar'' from the 1.3\,mm dust continuum emission. In the optically thin limit, the mass can be calculated as 
\begin{equation}
    M=R_{\rm dg}\frac{F_{\nu}D^2}{\kappa_\nu B_\nu(T)},
\end{equation}
where $R_{\rm dg}$ is the gas-to-dust ratio, $F_{\nu}$ is the flux density, $\kappa_{\nu}$ is the dust opacity per gram of dust, and $B_\nu(T)$ is the Planck function at the dust temperature of $T$. We use \textit{astrodendro} algorithm to extract the flux of the ``bar'' from the dendrogram which is an abstraction of the changing topology of the isosurfaces of a two-dimensional image as a function of contour levels \mycite{2008ApJ...679.1338R}. The algorithm organizes hierarchical structures into ``\textit{trees}'' consisting of ``\textit{branches}'', formed by combining ``\textit{leaves}'' in the terminology of \textit{astrodendro}. 
The following parameters are used in computing the dendrogram: the minimum pixel value \textit{min-value} is 3$\sigma_{\rm 1.3\,mm}$; the minimum difference in the peak intensity between neighboring structures \textit{min-delta} = 1$\sigma_{\rm 1.3\,mm}$; and the minimum number of pixels required for a structure to be considered an independent entity \textit{min-npix} $= N_{\rm beam}$, where $N_{\rm beam}$ is the number of pixels enclosed in the synthesized beam.  Fig.~\ref{cont_dendro} shows the identified ``bar'' as a tree and the potential protostars as leaves. We subtract the fluxes of two luminous leaves from the total flux of the tree to obtain the flux of the extended structure. 

Adopting a $R_{\rm dg}$ of 100, $\kappa_\nu$ of $1\,{\rm cm^{2}~g^{-1}}$ \mycite{1994A&A...291..943O}(interpolated to 1.3\,mm with the dust opacity index $\beta$ of 1.6), and a temperature of 20\,K, the mass of the ``bar'' is 6.2\,$M_{\odot}$. Besides the rotation, there is also a global velocity gradient of $\rm\sim41\,km~s^{-1}~pc^{-1}$ along the ``bar'' as seen in DCN (fig.~S\ref{pvs}). The infall rate along the ``bar'' can be estimated using Eq.~\ref{eq:3} which is 2.6$\times10^{-4}\,M_{\odot}~{\rm yr}^{-1}$, similar to the infall rate of the spiral system.

At the disk scale, assuming the $^{12}$CO outflow momentum is conserved from the protostellar wind momentum, we infer the accretion rate $\dot{M}_{\rm acc}$ from the mass-loss rate $\dot{M}_{\rm w}$ estimated from $^{12}$CO since $\dot{M}_{\rm acc}$ is proportional to $\dot{M}_{\rm w}$ via $\dot{M}_{\rm acc}=f\times\dot{M}_{\rm w}$ \mycite{2013A&A...551A...5E, 2016ARA&A..54..491B} where $f$ is the fraction between the accretion rate and mass-loss rate. We estimate  the gas mass of the outflow $m_{\rm i}$ at each channel following \mycite{2009ApJ...696...66Q}:
\begin{equation}
    m_{i}=1.39\times10^{-6}\exp\left({\frac{16.59}{T_{\rm ex}}}\right)(T_{\rm ex}+0.92)D^2\int S_{i}d\Omega,
\label{eq:12co}
\end{equation}
where $T_{ex}$, $D$, $S_{i}$ and $\Omega$ are the excitation temperature of $^{12}$CO (2-1), the distance in kpc, the flux in Jy of each pixel at channel $i$, and the solid angle, respectively. Eq.~\ref{eq:12co} is normalized to a $^{12}$CO-to-$\rm H_2$ abundance of $10^{-4}$ and a mean gas atomic weight of 1.36. We adopt a $T_{\rm ex}$ of 30\,K which is slightly higher than the 20\,K used above considering that the gas could be heated by the outflow. The velocity range is free of $^{13}$CO and C$^{18}$O (fig.~S\ref{co_spec}), thus we assume the $^{12}$CO emission is optically thin. Measuring the $^{12}$CO outflow projection length ($\lambda_{\rm max}$), we estimate the outflow dynamical timescale ($t_{\rm dyn}$), momentum $P_{\rm out}$, mechanical force $F_{\rm out}$, and thus mass-loss rate $\dot{M}_{\rm w}$ using:
\begin{equation}
    t_{\rm dyn}=\frac{\lambda_{\rm max}}{(v_{\rm max,b}+v_{\rm max,r})/2},
\end{equation}
\begin{equation}
    P_{\rm out}=\sum^{\rm blue} m_iv_i+\sum^{\rm red} m_iv_i,
\end{equation}
\begin{equation}
    F_{\rm out}=P_{\rm out}/t_{\rm dyn},
\end{equation}
\begin{equation}
    \dot{M}_{w}=\frac{F_{\rm out}}{v_w} \label{eq:masslossrate},
\end{equation}
Here $v_{\rm max,b}~{\rm and}~v_{\rm max,r}$ are the maximum blueshifted and redshifted $^{12}$CO velocities, respectively, and $v_{\rm w}$ is the wind velocity. We adopt a $v_{\rm w}$ of 500\,km s$^{-1}$ \mycite{2016ARA&A..54..491B, 1995ApJ...455..269L}. The outflow parameters are summarized in  table~S\ref{tab:outflow}. The mass-loss rates for the blue and red lobes are 5.7 and 3.6$\times 10^{-7}\,M_{\odot}~{\rm yr}^{-1}$, respectively, with a total of 9.3$\times 10^{-7}\,M_{\odot}~{\rm yr}^{-1}$. We adopt the fraction $f$ of 3 \mycite{1998ApJ...502L.163T}, resulting in an accretion rate of 2.8$\times10^{-6}\,M_{\odot}~{\rm yr}^{-1}$. 

\begin{table}[htbp]
\centering
\caption{Outflow parameters of I18134-1942}
\begin{tabular}{ccccccc}
\hline
\hline
Lobe & $\Delta v$      & $\lambda_{\rm max}$ & $t_{\rm dyn}$  & $P_{\rm out}$ & $F_{\rm out}$ & $\dot{M}_{\rm w}$                \\
     & km s$^{-1}$  & $10^{-3}$\,pc  & $10^3\,yr$ & $M_{\odot}~{\rm km~s^{-1}}$ & $10^{-2}~M_{\odot}~{\rm km~s^{-1}~yr^{-1}}$ & $10^{-5}\,M_{\odot}$ yr$^{-1}$ \\ \hline
Blue & {[}-30,0{]}                & 49.91          & 1.11       & 26.93 & 2.42 & 4.85                           \\
Red  & {[}20,50{]}                  & 18.41          & 0.41       & 4.74 & 1.16 & 2.31                           \\ \hline
\end{tabular}
\label{tab:outflow}
\end{table}

As we cannot derive the inclination of the outflow, the accretion rate has not been corrected for the projection with a factor of $\cos{i}/\sin^2{i}$. On a hemispherical surface, the probability distribution of the inclination angle $i$ is given by $P(i)\propto\cos{i}$, where $i$ ranges from 0 to $\pi$/2. The mean inclination angle of the outflow is calculated to be 
$$
    E(i)=\frac{\int_0^{\frac{\pi}{2}}i\cos i~{\rm d}i}{\int_0^{\frac{\pi}{2}}\cos i~{\rm d}i}=\frac{\pi}{2}-1\,{\rm radian}\approx32.7^{\circ}
$$
For the mean inclination angle, the very small (i.e. 5$^{\circ}$) and large (i.e. 85$^{\circ}$) inclination angle, the correction factors applied to the accretion rate are determined to be 2.9, 131.1, and 0.09, respectively. We estimated the mass inflow rate assuming the inclination of most of the inflows to be 45$^{\circ}$. Changing the inclination to 10$^{\circ}$ or 80$^{\circ}$, would increase or reduce the inflow rate by a factor of 5.7. The disk-mediated accretion rate inferred from the outflow would be comparable to the mass inflow rate only under the extreme cases.


\clearpage


\clearpage

\bibliography{18134na}{}


\vspace{1.5em}
\noindent\textbf{Acknowledgements}
X.M. is grateful to the financial support from China Scholarships Concil.
We acknowledge all the developers for their open-source codes and packages \textbf{Numpy \mycite{harris2020array}}, \textbf{Scipy \mycite{2020SciPy-NMeth}}, \textbf{Astropy \mycite{2013A&A...558A..33A}}, \textbf{Matplotlib \mycite{Hunter:2007}}, \textbf{Astrodendro \mycite{2019ascl.soft07016R}}, and \textbf{FERIA \mycite{2022PASP..134i4301O}}, which were crucial in this work.

\noindent\textbf{Funding}
T.L. acknowledges the supports by the National Key R\&D Program of China (No. 2022YFA1603100), National Natural Science Foundation of China (NSFC) through grants No.12073061 and No.12122307, and the Tianchi Talent Program of Xinjiang Uygur Autonomous Region.
B.Z. acknowledges support by the Natural Science Foundation of China (NSFC, grant No. U1831136 and U2031212) and Shanghai Astronomical Observatory (N-2020-06-09005).
X.M acknowledges financial support from the program of China Scholarships Council 202304910546.
G.G. and L.B. gratefully acknowledge support by the ANID BASAL project FB210003.
A.H. thanks the support by the S. N. Bose National Centre for Basic Sciences under the Department of Science and Technology, Govt. of India and the the CSIR-HRDG, Govt. of India for the funding of the fellowship.
X.L. acknowledges support from Strategic Priority Research Program of the Chinese Academy of Sciences Grant  XDB0800100.
PS was partially supported by a Grant-in-Aid for Scientific Research (KAKENHI Number JP23H01221) of JSPS. 
E.M. acknowledges funding from the Vilho, Yrjö ja Kalle Väisälän rahasto from the Finnish Academy of Science and Letters.
M.J., J.V., and D.T. acknowledge support from the Research Council of Finland grant 348342.
J.-E.L. was supported by the National Research Foundation of Korea (NRF) grant funded by the Korea government (MSIT) (grant numbers 2021R1A2C1011718 and RS-2024-00416859).
This research was carried out in part at the Jet Propulsion Laboratory, which is operated by the California Institute of Technology under a contract with the National Aeronautics and Space Administration (80NM0018D0004).
P.G., G.G. and L.Z. are supported by Chinese Academy of Sciences South America Center for Astronomy (CASSACA) Key Research Project E52H540201.
N.J.E thanks the Astronomy Department of the University of Texas for research support.

\vspace{0.5em}
\noindent\textbf{Author contributions}
Conceptualization: XM, TL, XL, NJE, JH, CWL, SZ, KT, SD, QZ, JEL, MJ, LZ, GG
Methodology: XM, QL, TL, XC, JH, KT, QZ, XL
Investigation: XM, TL, XL, XC, JH, CWL, SD, DY, SRD, QZ, JOC
Visualization: XM, TL, EM, SRD
Supervision: TL, GLW, CWL, MJ, XL
Writing--original draft: XM, TL, XL, KTK, QL, PFG, SZ, KT, SD
Writing--review \& editing: XM, TL, BZ, PFG, NJE, QZ, KTK, MJ, FX, WJ, HL, PS, GG, SQ, JMV, AT, ZR, SD, SL, QL, JH, PG, AH, YZ, JEL, SZ, EM, DT, LD, LB, PrG, XT, SRD, GW, CWL, JOC, YKZ, QG, KT, GLW, ZS
Validation: XM, QL, TL, ZR, SL, HL, JH, CWL, YKZ, XL, JOC, XC, PFG
Formal analysis: XM, QL, TL, XC, DY, JOC
Funding acquisition: TL, BZ, CWL, GLW, XL, LZ
Data curation: XM, TL, SZ, WJ, FX, LZ
Software: XM, TL, FX
Project administration: TL, GLW
Resources: CWL, BZ, GLW, DY, XL, TL


\vspace{0.2em}
\noindent\textbf{Competing interests} The authors declare they have no competing interests.

\noindent\textbf{Materials Availability}
All data needed to evaluate the conclusions in the paper are present in the paper and/or the Supplementary Materials. The ALMA data presented in this work can be found on ALMA Science Archive under project codes 2019.1.00685.S, 2021.1.00095.S, and 2021.1.00455.T.
The maser data is available via \url{https://cdsarc.cds.unistra.fr/viz-bin/cat/J/A+A/684/A86}.

\vspace{2.5em}
\noindent\large \textbf{Content for supplementary materials:}\\
Figures:\\
Fig.~S\ref{spiral_stack}: Stacked image of various gas tracers\\
Fig.~S\ref{cch_pvs}: The position-velocity diagrams of CCH. The elements in this figure are the same as in Fig.~\ref{h13cop_pvs}\\
Fig.~S\ref{pvs}: The PV diagrams of various molecules.\\
Fig.~S\ref{intensity_profile}: The intensity profiles of continuum emission and different molecular line emission\\
Fig.~S\ref{dcn_ch3oh}: The moment maps of DCN and CH$_3$OH\\
Fig.~S\ref{chi2}: The $\chi^2$ cube of the IRE fitting\\
Fig.~S\ref{ch3cn_mom_pv}: The moment maps and PV diagrams of CH$_3$CN J=12-11 K=3-5 transitions \\
Fig.~S\ref{orien_schem}: The schematic representation of the configuration of the envelope-disk system\\
Fig.~S\ref{outflow_sio_h2_maser}: The spatial distribution of the outflow and shocked gas\\
Fig.~S\ref{co_channel}: The channel maps of the blue lobe of CO outflow\\
Fig.~S\ref{spiral_model_continuum}: The spiral model with the highest Pearson correlation coefficient.\\
Fig.~S\ref{co_spec}: The spectra of CO, $\rm ^{13}CO$, and $\rm C^{18}O$\\

\vspace{0.5em}
\noindent Tables:\\
Table~S\ref{parameterrange}: The parameter range and the best-fit parameter set in the $\chi^2$ grid search\\
Table~S\ref{tab:outflow}: The outflow parameters\\
Table~S\ref{moltrans}: The molecule species and transitions used in this study

\clearpage

\clearpage

\setcounter{page}{1}

\setcounter{figure}{0}
\setcounter{table}{0}
\renewcommand{\figurename}{Supplementary Fig.}
\renewcommand{\figureautorefname}{Supplementary Fig.}
\renewcommand{\tablename}{Supplementary Table}
\renewcommand{\tableautorefname}{Supplementary Table}

\phantomsection
\label{sec:extendeddata}
\noindent{\Large \textbf{Supplementary Material}}

\vspace{1em}

\noindent\large \textbf{Content:}\\
Figures:\\
Fig.~S\ref{spiral_stack}: Stacked image of various gas tracers\\
Fig.~S\ref{cch_pvs}: The position-velocity diagrams of CCH. The elements in this figure are the same as in Fig.~\ref{h13cop_pvs}\\
Fig.~S\ref{pvs}: The PV diagrams of various molecules.\\
Fig.~S\ref{intensity_profile}: The intensity profiles of continuum emission and different molecular line emission\\
Fig.~S\ref{dcn_ch3oh}: The moment maps of DCN and CH$_3$OH\\
Fig.~S\ref{chi2}: The $\chi^2$ cube of the IRE fitting\\
Fig.~S\ref{ch3cn_mom_pv}: The moment maps and PV diagrams of CH$_3$CN J=12-11 K=3-5 transitions \\
Fig.~S\ref{orien_schem}: The schematic representation of the configuration of the envelope-disk system\\
Fig.~S\ref{outflow_sio_h2_maser}: The spatial distribution of the outflow and shocked gas\\
Fig.~S\ref{co_channel}: The channel maps of the blue lobe of CO outflow\\
Fig.~S\ref{spiral_model_continuum}: The spiral model with the highest Pearson correlation coefficient.\\
Fig.~S\ref{co_spec}: The spectra of CO, $\rm ^{13}CO$, and $\rm C^{18}O$\\

\vspace{0.5em}
\noindent Tables:\\
Table~S\ref{parameterrange}: The parameter range and the best-fit parameter set in the $\chi^2$ grid search\\
Table~S\ref{tab:outflow}: The outflow parameters\\
Table~S\ref{moltrans}: The molecule species and transitions used in this study

\clearpage
Supplementary Fig.~S\ref{spiral_stack} presents the stacked image of $\rm H^{13}CO^+$, CCH and 3\,mm continuum.

Supplementary Fig.~S\ref{cch_pvs} shows the PV diagrams of CCH along the selected paths.

Supplementary Fig.~S\ref{pvs} shows the PV diagrams for DCN, H$_2$CO, HC$_3$N, SO, CH$_3$OH and C$^{18}$O cut along the major axis of the ``bar". The gas kinematics can be decomposed into two modes: the elongated and velocity-coherent structure extending beyond 4$\arcsec$ and the diamond-shape structure with the clear spin-up feature. The velocity-coherent mode indicates the gas inflow along the ``bar'' while the ``diamond'' mode indicates the rotating and infalling motion at the center. The smooth transition between two modes may suggest that the gas is being transported to the central envelope and starts rotating.

Supplementary Fig.~S\ref{intensity_profile} presents the intensity profiles of continuum emission and different molecular line emission along the ``bar". The H$^{13}$CO$^+$ shows extended emission along the ``bar'' while CCH is less abundant at the center. Interestingly, despite the different angular resolution, the 3\,mm continuum dust and C$^{18}$O in 1.3\,mm observation show a similar extent, suggesting that C$^{18}$O is tracing a more flattened gas reservoir than DCN and 1.3\,mm continuum, which is central dominant (the ``nucleus'') with two shoulders and long tails (the ``bar'') on both sides in the profiles. 

Supplementary Fig.~S\ref{dcn_ch3oh} shows the zoomed-in Moment 0 and Moment 1 maps of DCN and CH$_3$OH. Close to the ``nucleus", both molecular lines show tails connected to larger-scale gas streamers, as outlined by cyan dashed lines.

Supplementary Fig.~S\ref{chi2} shows the $\chi^2$ cube of the IRE fitting.

Supplementary Fig.~S\ref{ch3cn_mom_pv} shows the moment maps and PV diagrams of CH$_3$CN J=12-11 K=3-5 transitions, which well reveal the geometry and kinematics of the pseudodisk.

Supplementary Fig.~S\ref{orien_schem} is a schematic representation of the inversion of the line-of-sight velocity field between the envelope and the disk.

Supplementary Fig.~S\ref{outflow_sio_h2_maser} shows the spatial distribution of CO outflow and shocked gas traced by SiO, $\rm H_2$, masers.

Supplementary Fig.~S\ref{co_channel} shows the channel maps of $^{12}$CO blueshifted high-velocity emission. The jet direction revealed in higher-velocity channels is remarkably different from the one seen in lower-velocity channels, indicating potential jet precession.

Supplementary Table ~S\ref{parameterrange} lists the parameter space of the $\chi^2$ search, and the best-fit parameter set with the corresponding lowest $\chi^2$ value.

Supplementary Table ~S\ref{tab:outflow} summarizes the calculated outflow parameters.

Supplementary Table~S\ref{moltrans} summarizes the molecule species and the transitions used in this study.

Supplementary Fig.~S\ref{spiral_model_continuum} presents the best spiral model on the 3 mm continuum emission data. 

Supplementary Fig.~S\ref{co_spec} shows the spectra of $^{12}$CO, $^{13}$CO, and C$^{18}$O, and the selected velocity ranges for outflow.

\begin{figure*}[h]
\centering
\includegraphics[width=0.99\linewidth]{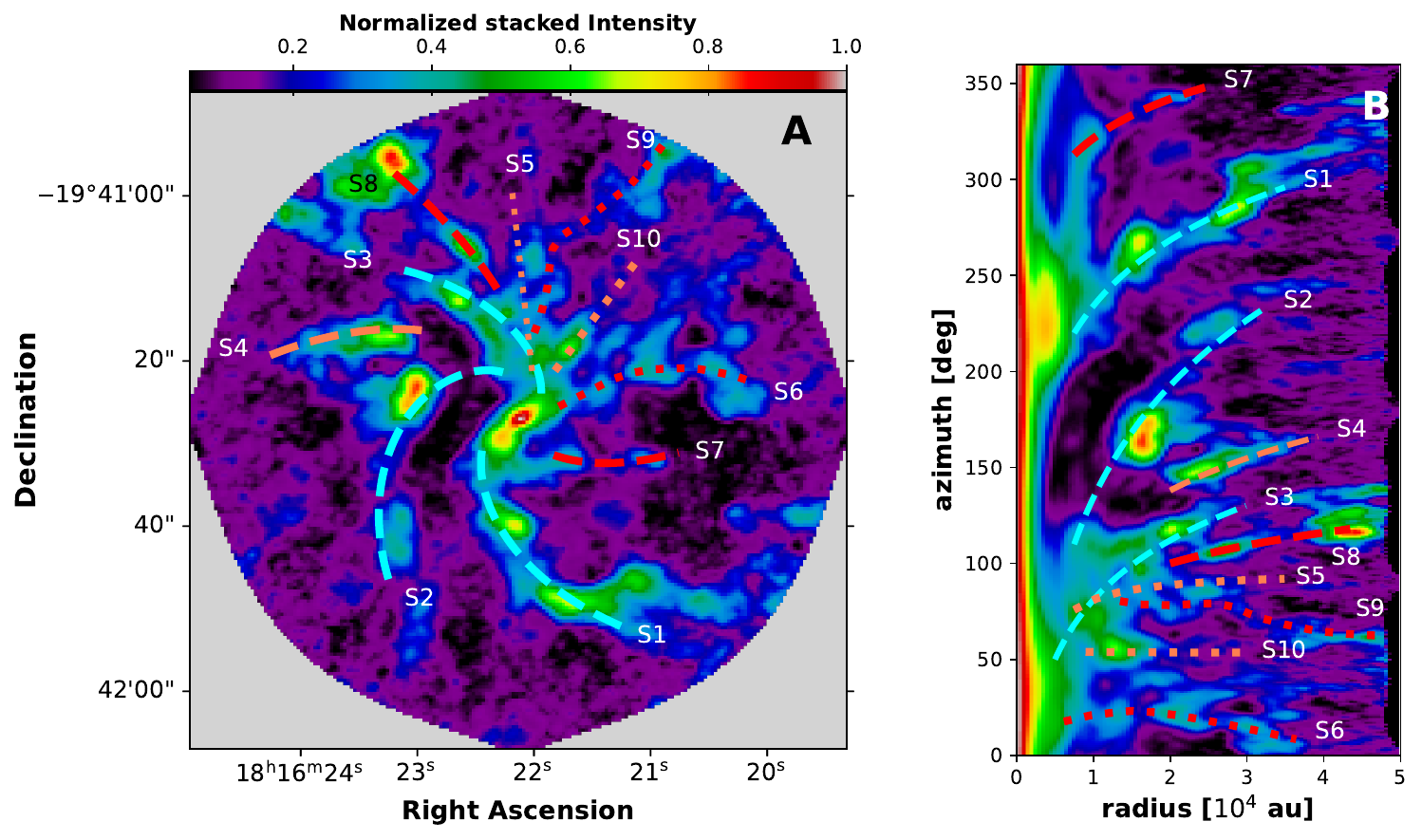}
\caption{\textbf{The overall gas distribution reveals an intricate spiral-like hub-filament system.} \textbf{Panel A} shows the stacked image of $\rm H^{13}CO^+$, CCH and 3\,mm continuum after normalization while \textbf{panel B} shows the polar projection of \textbf{panel A}. The cyan, orange, and red curves outline the filaments identified in 3\,mm continuum, $\rm H^{13}CO^+$, and CCH in Fig.~\ref{b3cont_spiral} and \ref{spiral_h13cop}. }
\label{spiral_stack}
\end{figure*}

\begin{figure*}[h]
\centering
\includegraphics[width=0.99\linewidth]{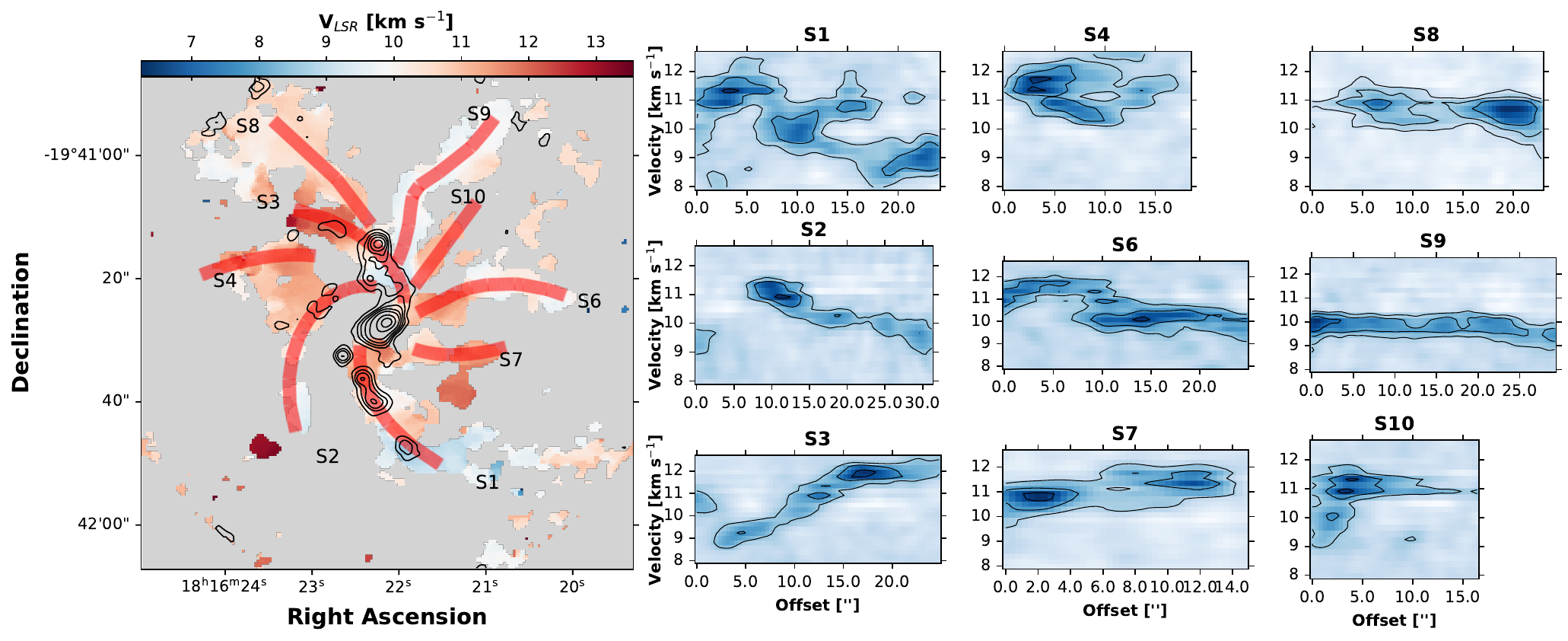}
\caption{\textbf{The velocity coherent filaments traced by CCH.} Same as in Fig.~\ref{h13cop_pvs}. } 
\label{cch_pvs}
\end{figure*}


\begin{figure}[h]
\centering
\includegraphics[width=1\linewidth]{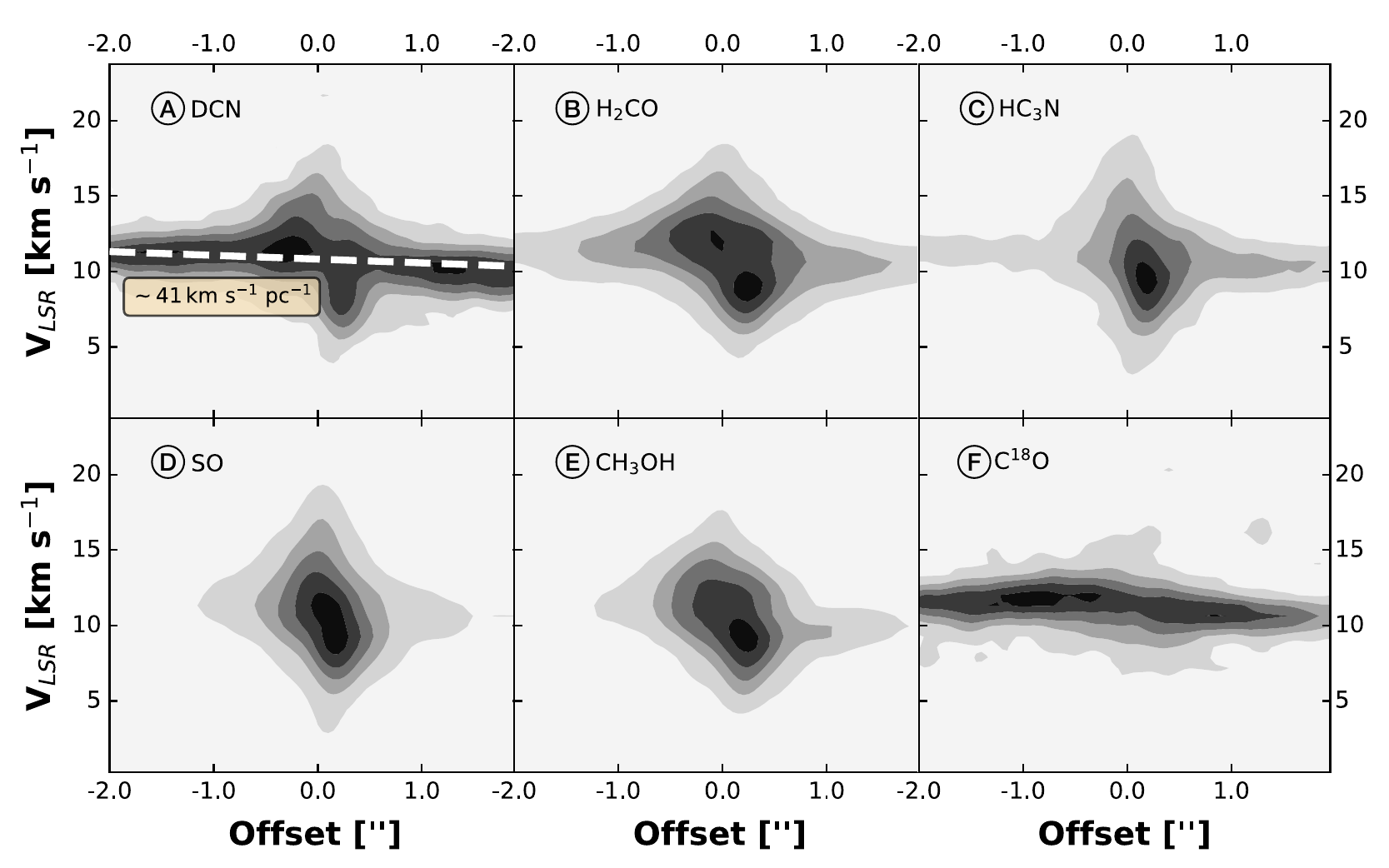}
\caption{\textbf{The kinematics of the streamer-envelope system.} The position-velocity diagrams of different molecules (marked in the upper left corner of each panel) along the same cut are plotted as filled contours. All the contours start from 10$\%$ of the peak intensity and increase on 20$\%$ steps. The direction of the PV cut is labeled as the green dashed line in Fig.~\ref{dcn_con_streamer}b but for the inner 4$\arcsec$. The white dashed lines in panel \textbf{A} represent the global velocity gradient along the ``bar'' of $\rm\sim41\,km~s^{-1}~pc^{-1}$.
}
\label{pvs}
\end{figure}

\begin{figure}[h]
\centering
\includegraphics[width=0.8\linewidth]{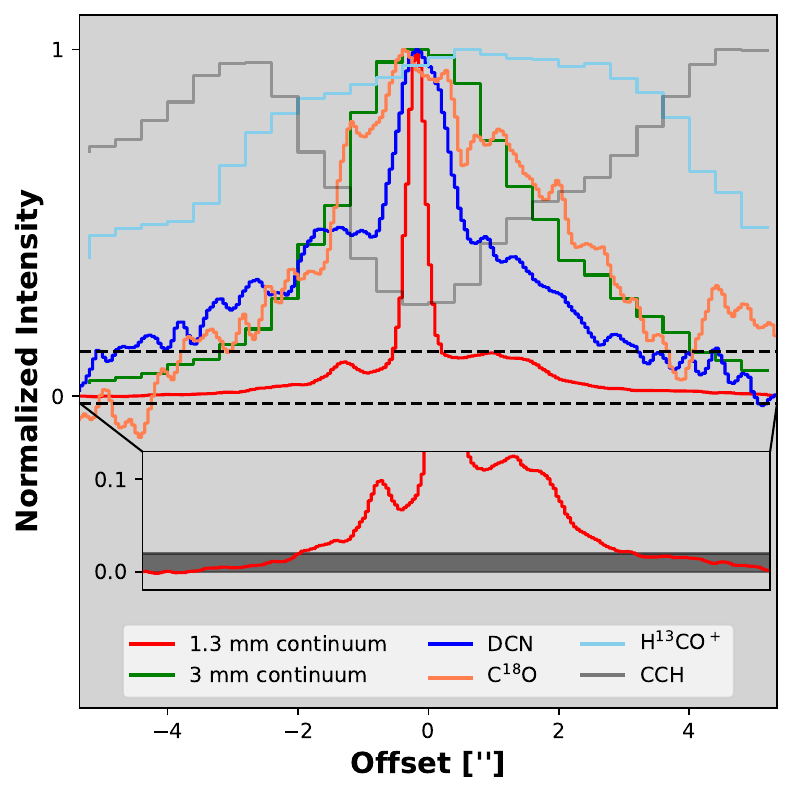}
\caption{\textbf{The normalized intensity profile of the continuum maps and different molecules along the major axis of the bar}. The dashed horizontal lines mark the zoom-in range. The grey band in the zoom-in axis represents the 3-$\sigma_{1.3\,mm}$.}
\label{intensity_profile}
\end{figure}

\begin{figure}[h]
\centering
\includegraphics[width=0.85\linewidth]{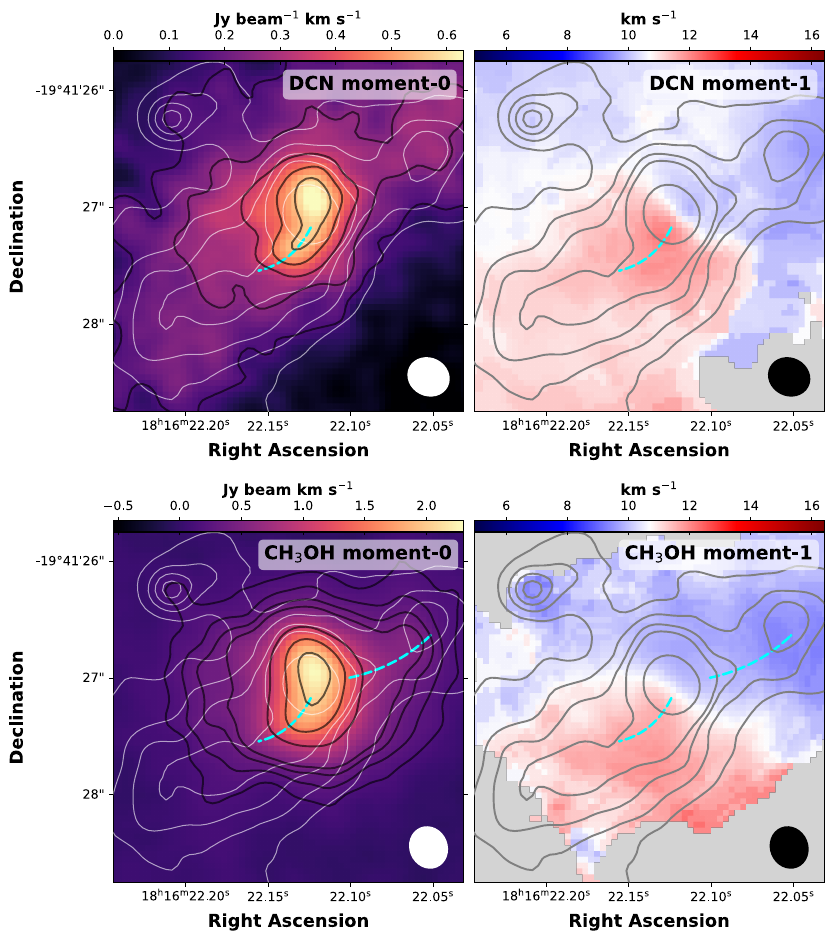}
\caption{ \textbf{The integrated intensity and intensity-weighted velocity maps of DCN and $\rm CH_3OH$.} This figure shows the zoom-in view of the integrated intensity (\textbf{left}) and centroid velocity (\textbf{right}) of DCN (\textbf{top}) and CH$_3$OH (\textbf{bottom}). The 1.3\,mm continuum emission is shown as the white contours in the left column and grey contours in the right column. The black contours in the upper left panel are the integrated intensity of DCN at [3, 5, 7, 9, 11]$\times 0.05$ Jy beam$^{-1}$ km s$^{-1}$. The black contours in the lower left panel are the integrated intensity of CH$_3$OH at [3, 5, 7, 9, 20, 30, 40]$\times 0.05$ Jy beam$^{-1}$ km s$^{-1}$. The cyan dashed curves trace the `tails' ---- the terminal end of the converging streamers. The synthesized beam is shown at the lower right corners.
}
\label{dcn_ch3oh}
\end{figure}

\begin{figure*}[h]
\centering
\includegraphics[width=1\linewidth]{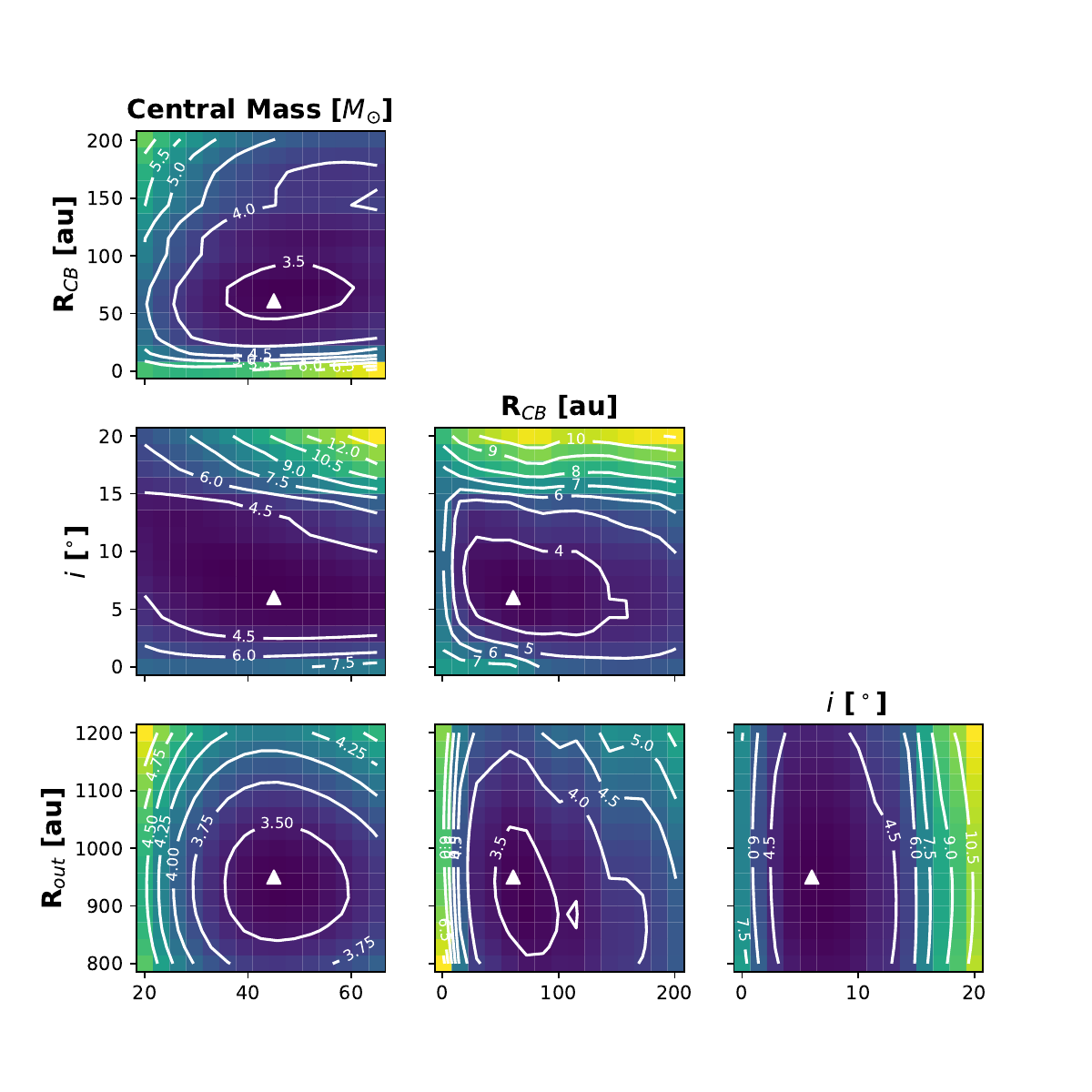}
\caption{ \textbf{$\chi^2$ grid search performed in $\rm CH_3OH~4_2-3_1~E1~vt=0$}. These colormaps represent the slices of the $\chi^2$ cube across the best-fit parameter set marked by the white triangles. The parameter ranges are summarized in  table~S\ref{parameterrange}.   }
\label{chi2}
\end{figure*}

\begin{figure}[h]
\centering
\includegraphics[width=0.7\linewidth]{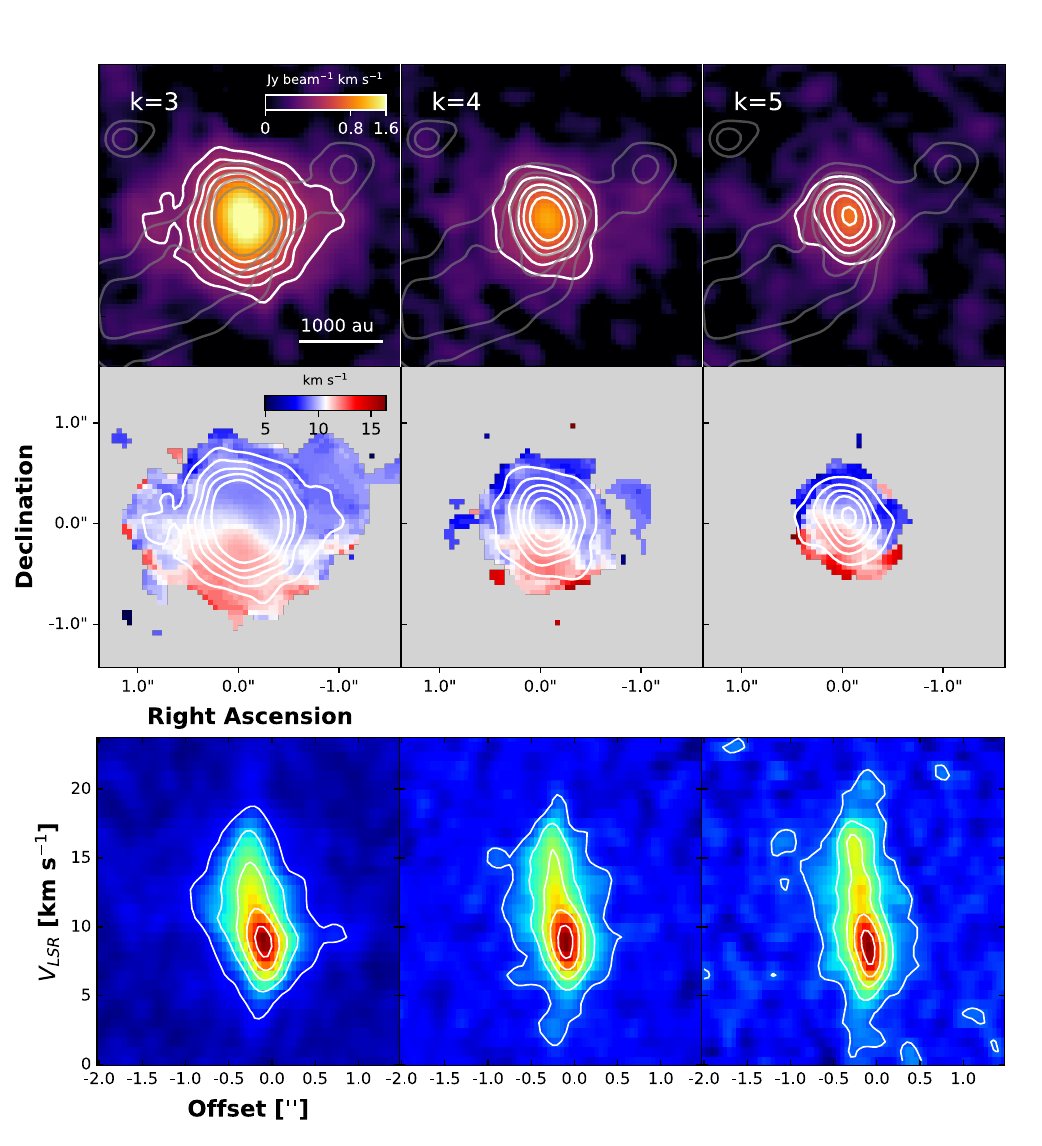}
\caption{
\textbf{Moment maps and PV diagrams of CH$_3$CN J=12-11 K=3-5. } 
The top and middle rows are their integrated intensity maps and centroid velocity maps, respectively, and the bottom row is their PV diagrams. The white contours in the top and middle rows are at [3,5,7,9,12]$\times$0.06\,${\rm Jy~beam^{-1}~km~s^{-1}}$. The gray contours in the top rows are 1.3\,mm continuum emission at [10,15,20,30]$\times\sigma_{\rm 1.3\,mm}$. The white contours in the bottom rows are at 10\%, 30\%, 50\%, 70\%, 90\% of the peak intensities.
}
\label{ch3cn_mom_pv}
\end{figure}

\begin{figure}[h]
\centering
\includegraphics[width=0.8\linewidth]{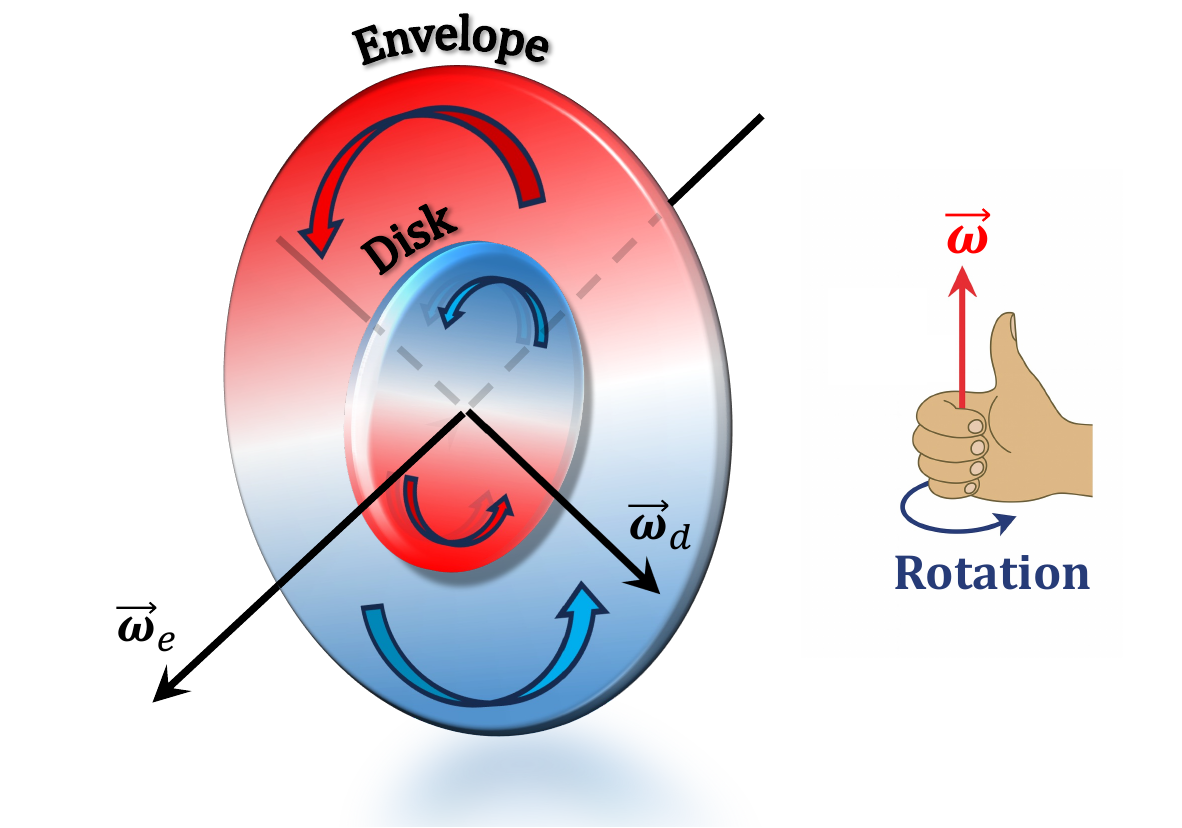}
\caption{\textbf{Schematic representation of inclination difference of the envelope-disk system.} Vectors $\vec{\omega}_e$ and $\vec{\omega}_d$ denote the angular momentum vectors of the envelope and the disk, respectively. The rotation of both the envelope and disk follows the right-hand rule.   }
\label{orien_schem}
\end{figure}

\begin{figure*}[h]
\centering
\includegraphics[width=0.9\linewidth]{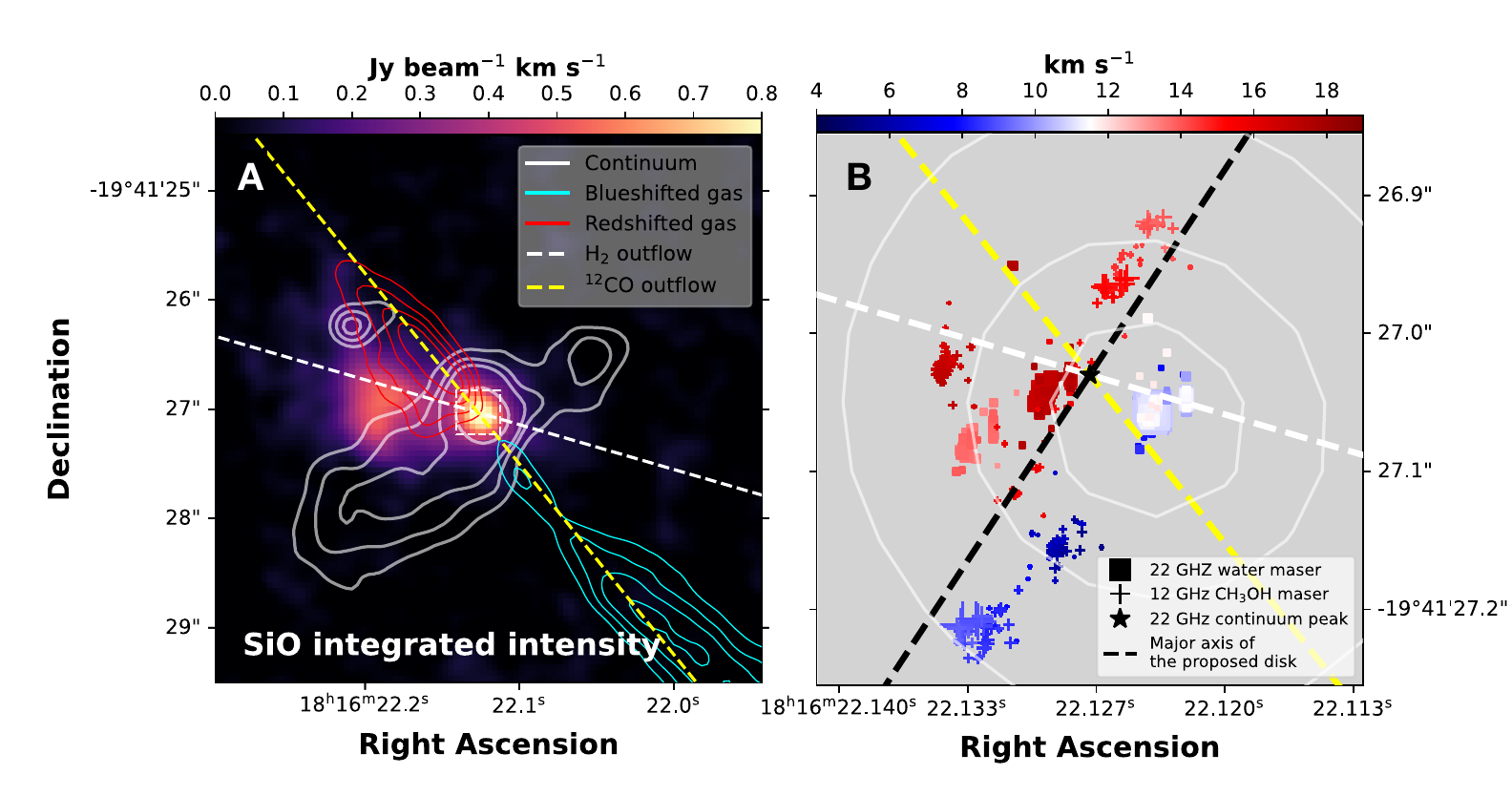}
\caption{
\textbf{Misalignment between outflow gas traced by $^{12}$CO and the shocked gas traced by SiO, $\rm H_2$ and 22 GHz water masers.} \textbf{A,} the background shows the SiO integrated intensity. The reshifted and blueshifted outflow gas traced by $^{12}$CO are shown as red and cyan contours respectively, and the white contours are the 1.3\,mm continuum emission. The red and cyan contours are at the level of [7, 9, 11, 13] $\times\sigma_{^{12}\rm CO}$ where $\rm \sigma_{^{12}CO}=0.12\,Jy~beam^{-1}~km~s^{-1}$. Contours of the continuum emission are at the level of [15, 19, 30, 50, 80, 150]$\times \sigma_{1.3\,\rm mm}$ to discard the diffuse emission and emphasize the dusty streamer. The white dashed line represents the direction of $\rm H_2$ knots \mycite{10.1046/j.1365-8711.2003.06419.x,2024A&A...684A..86B}. The dashed square at the center of the image shows the view of panel B.
\textbf{B,} Maser distribution inside the innermost 400 au region. The square and plus symbols mark the 22 GHz water masers and the 12 GHz methanol masers, respectively. The white dashed line is the same as in panel a. The black dashed line represents the major axis of the proposed disk.
}
\label{outflow_sio_h2_maser}
\end{figure*}

\begin{figure}[h]
\centering
\includegraphics[width=1\linewidth]{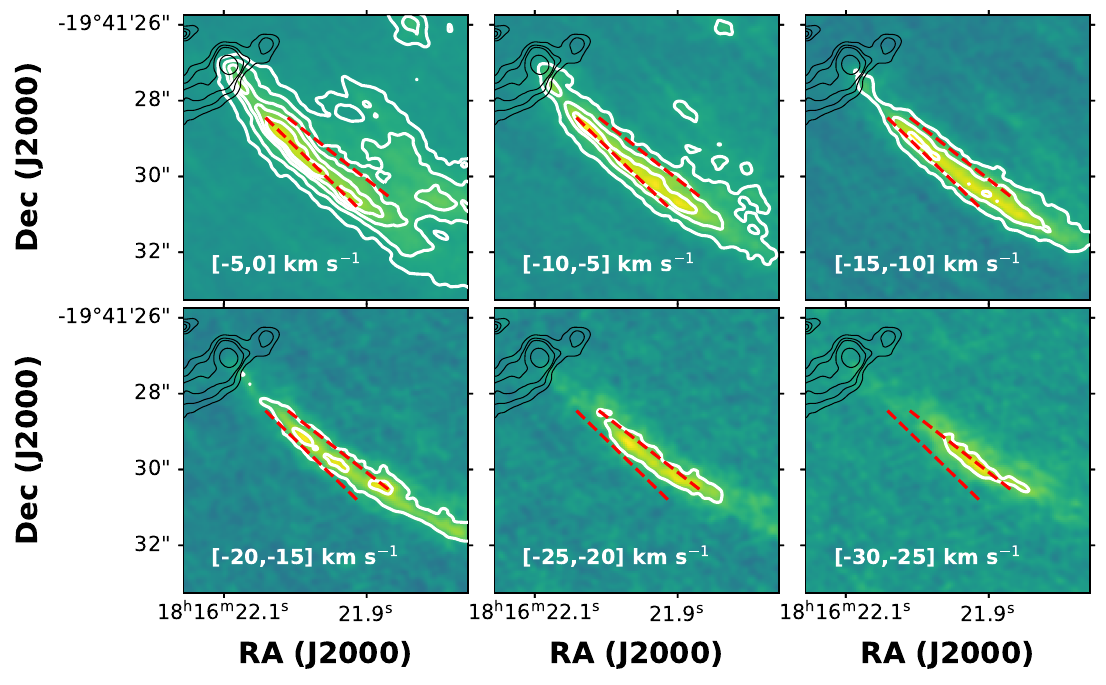}
\caption{\textbf{The $^{12}$CO integrated intensity maps of the blue lobe over every 5\,km s$^{-1}$ from -30 to 0\, km s$^{-1}$.} The background and white contours are for the $^{12}$CO integrated intensity at [10, 20, 30, 40, 50]$\times\sigma_{^{12}\rm CO}$, where $\sigma_{^{12}\rm CO}=12\,{\rm mJy~beam^{-1}~km~s^{-1}}$. The black contours mark the 1.3\,mm continuum emission at [9, 13, 17, 50]$\times$$\sigma_{1.3\,mm}$. The two red dashed lines outline the jet direction in low- and high-velocity channels. }
\label{co_channel}
\end{figure}

\begin{figure}[h]
\centering
\includegraphics[width=0.8\linewidth]{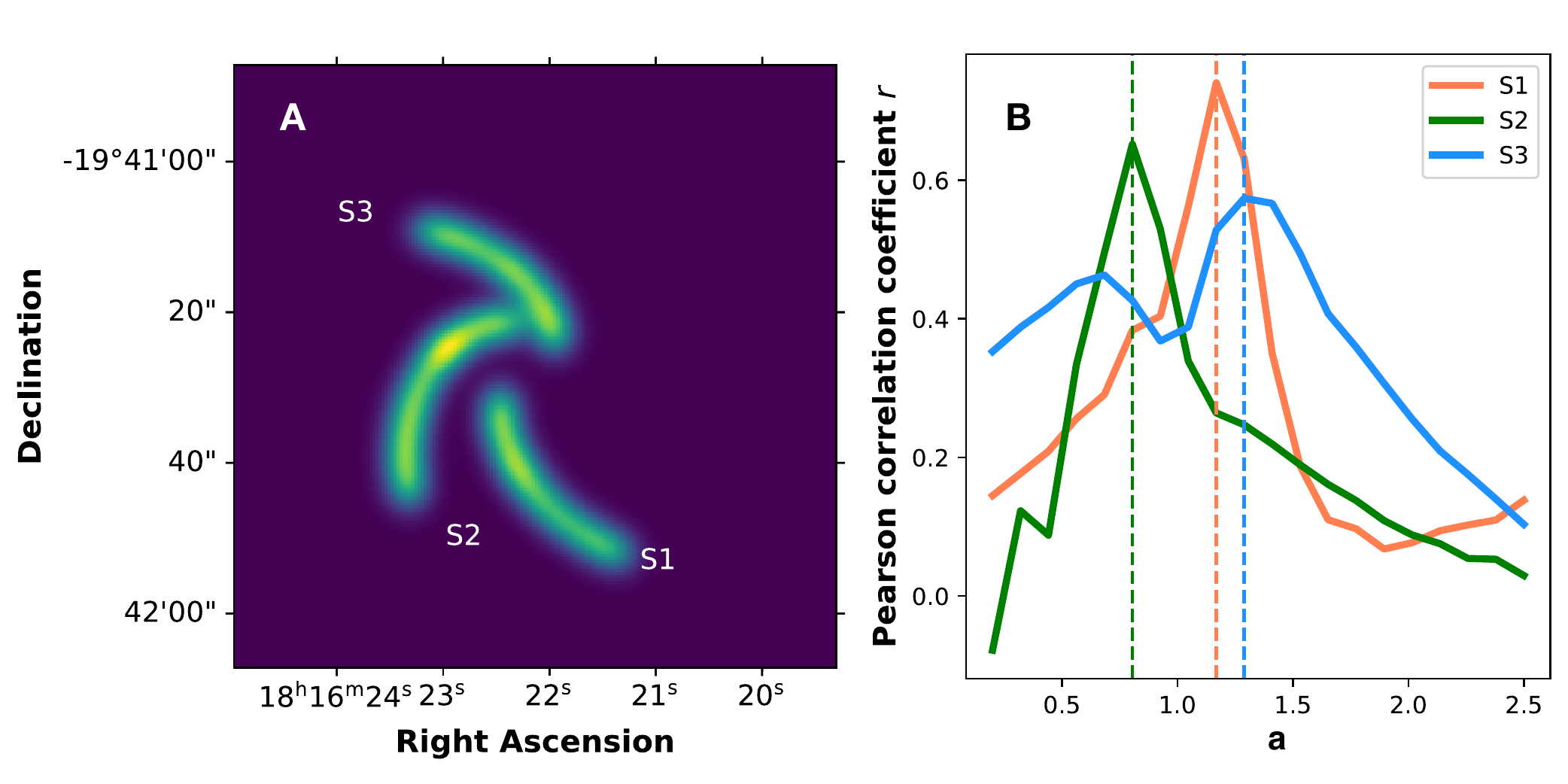}
\caption{\textbf{A}, the best-match spiral model. \textbf{B}, the Pearson correlation coefficient $r$ of the growth rate $a$ ranging from 0.2 to 2.5.}
\label{spiral_model_continuum}
\end{figure}

\begin{figure}[h]
\centering
\includegraphics[width=0.8\linewidth]{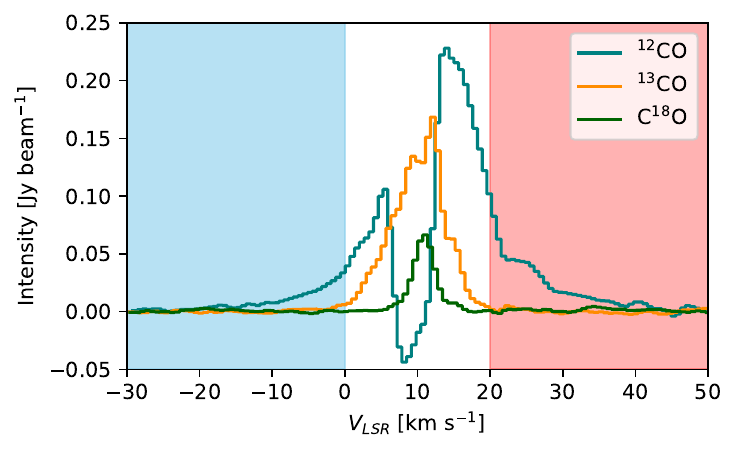}
\caption{\textbf{The spectra of $^{12}$CO, $^{13}$CO, and C$^{18}$O at the 1.3\,mm continuum peak position.} Different molecules are labeled as in the legend. The blue-shaded and red-shaded areas are the selected velocity ranges for blueshifted and redshifted lobes. }
\label{co_spec}
\end{figure}

\clearpage

\begin{table*}[h]
\centering
\caption{Free parameters and the best-fit parameter set in the $\chi^2$ grid search for CH$_3$OH}
\begin{tabular}{lccc}
\toprule
\toprule
Parameters                 & Range         & Best value             & Reduced $\chi^2$             \\
\midrule
$M \ [M_\odot]$                      & 20--70                 & 45                      & 3.4                        \\
$i$ $\ [^\circ]$                       & 0--20                  & 6                       &                            \\
$r_{\text{CB}} \ [{\rm au}]$               & 0--200                 & 60                         &                            \\
$r_{\text{out}} \ [{\rm au}]$              & 800--1200              & 950                     &                            \\

\bottomrule
\end{tabular}
\label{parameterrange}
\end{table*}

\begin{table}[htbp]
\centering
\caption{Outflow parameters of I18134-1942}
\begin{tabular}{ccccccc}
\hline
\hline
Lobe & $\Delta v$      & $\lambda_{\rm max}$ & $t_{\rm dyn}$  & $P_{\rm out}$ & $F_{\rm out}$ & $\dot{M}_{\rm w}$                \\
     & km s$^{-1}$  & $10^{-3}$\,pc  & $10^3\,yr$ & 10$^{-1}\,M_{\odot}~{\rm km~s^{-1}}$ & $10^{-3}~M_{\odot}~{\rm km~s^{-1}~yr^{-1}}$ & $10^{-7}\,M_{\odot}$ yr$^{-1}$ \\ \hline
Blue & {[}-30,0{]}                & 49.91          & 1.11       & 3.16 & 2.85 & 5.70                           \\
Red  & {[}20,50{]}                  & 18.41          & 0.41       & 0.73 & 1.79 & 3.59                           \\ \hline
\end{tabular}
\label{tab:outflow}
\end{table}

\begin{table}[H]
\centering
\caption{The molecule species and transitions used in this study}
\begin{tabular}{llll}
\hline
\hline
Species          &  & Rest Frequency (GHz)                 & $E_u$ (K) \\ \hline
3\,mm            &                      &                               &           \\
$\rm H^{13}CO^+$ & $J=1-0$                         & 86.754288          & 4.16      \\
CCH              & $N_{J,K}=1_{3/2,2}-0_{1/2,1}$   & 87.316898          & 4.19      \\ \hline
1.3\,mm          &                                 &                    &           \\
SiO     &   $J=5-4$ & 217.10498 &   31.26\\
$\rm H_2CO$      & 3(0, 3)-2(0, 2)   & 218.222192         & 20.96     \\
DCN              & $J=3-2$                         &   217.238538       & 20.85     \\
$\rm HC_3N$      & $J=24-23$                       &  218.324723        & 130.98    \\
$\rm CH_3OH$     & $4_2-3_1~E1~vt=0$               &   218.440063       & 45.46     \\
$\rm CH_3CN$     & $J=12-11, K=3$                   &   220.7090165       & 133.16    \\
                 & $J=12-11, K=4$                   &   220.6792869       & 183.15     \\
                 & $J=12-11, K=5$                   &   220.6410839      & 247.40    \\
C$^{18}$O        & $J=2-1$                         & 219.560354         & 15.81     \\
SO               & 6(5)-5(4)                       & 219.949442         & 34.98     \\
$^{12}$CO        & $J=2-1$                         & 230.538            & 16.59     \\ \hline
\end{tabular}
\label{moltrans}
\end{table}


\end{document}